\documentclass[letter,apj]{emulateapj-rtx4}
\pdfoutput=1
\usepackage{graphicx}
\usepackage{enumerate}
\usepackage{amssymb, amsmath, amsfonts}
\usepackage{gensymb}
\usepackage{mathtools}
\usepackage{natbib}
\usepackage{color}
\usepackage{hyperref}
\usepackage{ulem}
\usepackage{soul}
\usepackage{url}
\newcommand{\Hipparcos}{{\sl Hipparcos}}
\newcommand{\Gaia}{{\sl Gaia}}
\newcommand{\HST}{{\sl HST}}

\newcommand{\Msun}{\mbox{$M_{\sun}$}}

\newcommand{\Lsun}{\mbox{$L_{\sun}$}}

\newcommand{\Mjup}{\mbox{$M_{\rm Jup}$}}
\newcommand{\Rjup}{\mbox{$R_{\rm Jup}$}}
\newcommand{\Lbol}{\mbox{$L_{\rm bol}$}}

\newcommand{\Teff}{\mbox{$T_{\rm eff}$}}
\newcommand{\logg}{\mbox{$\log(g)$}}

\begin{document}

\altaffiltext{1}{Department of Physics, University of California, Santa Barbara, Santa Barbara, CA 93106, USA}
\altaffiltext{2}{Gemini Observatory, Northern Operations Center, 670 N.~Aohoku Place, Hilo, HI 96720, USA}
\altaffiltext{3}{Institute for Astronomy, University of Edinburgh, Royal Observatory, Blackford Hill, Edinburgh, EH9 3HJ, UK}
\altaffiltext{4}{Department of Astronomy, The University of Texas at Austin, Austin, TX 78712, USA}
\altaffiltext{5}{American Museum of Natural History, New York, NY}
\altaffiltext{6}{Science Support Office, Directorate of Science, European Space Research and Technology Centre (ESA/ESTEC), Keplerlaan 1, 2201 AZ Noordwijk, The Netherlands}

\title{A Dynamical Mass of $70 \pm 5~M_{\rm Jup}$ for Gliese 229B, the First T Dwarf}
\author{
Timothy D.~Brandt\altaffilmark{1}, 
Trent J.~Dupuy\altaffilmark{2,3}, 
Brendan P.~Bowler\altaffilmark{4},
Daniella C.~Bardalez Gagliuffi\altaffilmark{5},
Jacqueline Faherty\altaffilmark{5},
G.~Mirek Brandt\altaffilmark{1}, and
Daniel Michalik\altaffilmark{6}
}

\begin{abstract}
We combine Keck/HIRES radial velocities, imaging with HiCIAO/Subaru and the {\it Hubble Space Telescope}, and absolute astrometry from {\it Hipparcos} and {\it Gaia} to measure a dynamical mass of $70 \pm 5$~\Mjup\ for the brown dwarf companion to Gl~229.  Gl~229B was the first imaged brown dwarf to show clear signs of methane in its atmosphere.  Cooling models have been used to estimate a mass in the range of 20--55~\Mjup, much lower than our measured value.  We argue that our high dynamical mass is unlikely to be due to perturbations from additional unseen companions or to Gl~229B being itself a binary, and we find no evidence of a previously claimed radial velocity planet around Gl~229A.  Future {\it Gaia} data releases will confirm the reliability of the absolute astrometry, though the data pass all quality checks in both {\it Hipparcos} and {\it Gaia}.  Our dynamical mass implies a very old age for Gl~229, in some tension with kinematic and activity age indicators, and/or shortcomings in brown dwarf cooling models.  Gl~229B joins a small but growing list of T dwarfs with masses approaching the minimum mass for core hydrogen ignition. 
\end{abstract}

\maketitle

\section{Introduction} \label{sec:intro}

Brown dwarfs' physical and atmospheric properties bridge giant planets to stars \citep{Burrows+Marley+Hubbard+etal_1997}. They can have masses of up to $\sim$80~$M_{\rm Jup}$, depending on metallicity \citep[e.g.,][]{Burrows+Hubbard+Lunine+etal_2001,2019ApJ...879...94F}, as degeneracy pressure prevents the core from reaching a temperature necessary to sustain hydrogen fusion. This lack of an internal energy-generating mechanism causes substellar objects to continuously contract, cool, and dim: objects of very different masses can have the same luminosity and temperature if they were born at different times. To test models of brown dwarfs, we need to know both mass {\it and} age 
together with other observable properties like luminosity.

Although $\sim$10$^3$ brown dwarfs are now known \citep[e.g.,][]{Smart+Marocco+Caballero+etal_2017}, only $\sim$20 brown dwarf binary systems have fully determined orbits and dynamical mass measurements for their individual components \citep{Dupuy+Liu_2017}. Such dynamical masses have typically relied on a combination of high-angular resolution imaging to measure relative astrometry and flux ratios and absolute astrometry from wide-field imaging to measure the parallax and photocenter orbit. While current adaptive optics (AO) imaging techniques can resolve most brown dwarf binaries down to separations of $\sim$1\,AU, such systems are very rare, and in practice dynamical masses have more typically relied on 2--4\,AU binaries. The correspondingly long orbital periods of such systems, 10--30~years, have required a major effort spanning over a decade of patient orbit monitoring. And in the very small subset of cases where brown dwarf binaries orbit solar-type stars, the stellar age and metallicity have provided the strongest tests of substellar evolutionary models \citep[e.g.,][]{2009ApJ...692..729D,2014ApJ...790..133D,Crepp+Johnson+Fischer+etal_2012,Cardoso_2012}.

The first dynamical mass measurement of the components of an L/T transition binary showed that some of the most widely used models \citep[e.g.,][]{Baraffe+Chabrier+Barman+etal_2003} fail to reproduce coevality \citep{Dupuy+Liu+Leggett+etal_2015}. Recent age estimates and dynamical mass measurements of the old ultracool dwarfs Gl~758B, HD~4747B, and HR~7672B, however, showed generally good agreement with models \citep{Brandt+Dupuy+Bowler_2018} and resolved some earlier discrepancies \citep{Bowler+Dupuy+Endl+etal_2018,Calissendorff+Janson_2018}.  

Like the first T-dwarf binary to yield a dynamical mass \citep{2008ApJ...689..436L}, most T~dwarfs have measured masses that are relatively low (30--50\,\Mjup), as expected given their low temperatures (950--1250\,K) and the typical ages of field dwarfs (1--5\,Gyr). There have been some notable exceptions, with measured T~dwarf masses at the
very high end of, and possibly at odds with, expectations.
WISE~J0720$-$0846B (T5.5, $66\pm4$\,\Mjup; \citealp{Dupuy+Liu+Best+etal_2019}) and Gl~758B ($\sim$T8, $38.1_{-1.5}^{+1.7}$\,\Mjup; \citealp{Vigan+Bonnefoy+Ginski+etal_2016,Brandt+Dupuy+Bowler_2018}) both require ages of several Gyr, according to evolutionary models, in order to have masses as high as observed given their luminosities. According to \citet{Cardoso_2012}, the case of $\epsilon$~Ind~BC is similar, with masses of $68.0\pm0.9$\,\Mjup\ and $53.1\pm0.3$\,\Mjup\ for the T1.5 and T6 components \citep{Kasper+Burrows+Brandner_2009,King+McCaughrean+Homeier+etal_2010}. However, \cite{Dieterich+Weinberger+Boss+etal_2018} find the masses to be much higher, $75.0\pm0.8$\,\Mjup\ and $70.1\pm0.7$\,\Mjup.  This would present a severe challenge to most evolutionary models, which predict that such massive objects cannot cool sufficiently to display methane absorption.
Another highly problematic case for models is the late-T dwarf HD~4113C. Its dynamical mass of $66^{+5}_{-4}$~$M_{\rm Jup}$ \citep{Cheetham+Segransan+Peretti+etal_2018} is significantly higher than the maximum mass of $\approx$40\,\Mjup\ predicted from models for such a low luminosity object ($\approx10^{-6}$\,\Lsun).

Anomalously high masses could always be due to unresolved multiplicity.  Ruling out this scenario can be difficult, especially at older ages where a lower-mass companion would contribute little to the integrated spectrum or colors.
A larger sample of dynamical masses for well-characterized objects is needed to determine whether discrepancies with models are due to shortcomings in the models themselves, unresolved multiplicity or to other observational biases.

In this paper we present a precise dynamical mass measurement for the T7~dwarf Gl~229B, the very first object of its type (and one of the first brown dwarfs ever) to be discovered \citep{Nakajima+Oppenheimer+Kulkarni+etal_1995,Oppenheimer+Kulkarni+Matthews+etal_1995}.  
We use the {\sl Hipparcos}--{\sl Gaia} Catalog of Accelerations (HGCA;  \citealt{Brandt_2018}), combined with relative astrometry from direct imaging and 17 years of precision radial velocities, to precisely determine its mass and constrain its orbital parameters. We find that this object joins the short, yet growing, list of ``massive'' T dwarfs.

\section{System Age and Metallicity} \label{sec:age_metallicity}

Some of the earliest age analyses 
of Gl~229 
accompanied 
the discovery of its brown dwarf companion. 
\cite{Nakajima+Oppenheimer+Kulkarni+etal_1995} noted that the star's kinematics are cold, consistent with
a lack of dynamical heating due to youth.  However, unlike young and active low-mass stars, Gl~229 shows H$\alpha$ in absorption rather than emission. \cite{Nakajima+Oppenheimer+Kulkarni+etal_1995} argued that Gl~229's observed H$\alpha$ absorption rules out very young ages ($\lesssim$500~Myr) and adopted the wide range of 0.5--5~Gyr. 
\cite{Leggett+Toomey+Geballe+etal_1999} derived a consistent age of 0.5--1~Gyr and a mass of 25--35~\Mjup~for Gl~229B by fitting the brown dwarf's colors and luminosity to evolutionary models.
However, \cite{Leggett+Hauschildt+Allard+etal_2002} derived a younger age of $\sim$30~Myr and a metallicity of $[\rm {M/H}] \approx -0.5$\,dex by simultaneously fitting low-resolution spectra of both Gl~229A and B with AMES-Cond models \citep{Allard+Hauschildt+Alexander+etal_2001}. 

Gl~229A is an M1V dwarf \citep{Kirkpatrick+Henry+McCarthy_1991,Cushing+Roellig+Marley+etal_2006}, a class of star 
for which age is difficult to determine
\citep{West+Hawley+Bochanski+etal_2008}.  Nevertheless, we can use the star's rotation, activity, and kinematics to provide some constraints on the age.  We revisit the age of the system here using only the properties of the primary star, Gl~229A.  We also review literature measurements on the star's composition, particularly given the fit by \cite{Leggett+Hauschildt+Allard+etal_2002} at a strongly subsolar metallicity.

\subsection{Metallicity} \label{subsec:metallicity}

Recent measurements of Gl~229A's metallicity from medium- and high-resolution spectra yield approximately solar values.  \cite{Gaidos+Mann_2014} derived a spectroscopic metallicity of $[{\rm Fe/H}] = 0.12 \pm 0.10$\,dex from medium-resolution SpeX spectroscopy across the $JHK$ bands. \cite{Gaidos+Mann+Lepine+etal_2014} used high signal-to-noise ratio, medium-resolution optical spectra calibrated to M dwarfs using wide binaries to obtain $[{\rm Fe/H}] = 0.02 \pm 0.11$\,dex, consistent with the previous measurement.

\cite{Neves+Bonfils+Santos+etal_2013} published two metallicities from HARPS spectra: $[{\rm Fe/H}] = 0.11$\,dex and $[{\rm Fe/H}] = -0.05$\,dex using the \cite{Johnson+Apps_2009} and \cite{Neves+Bonfils+Santos+etal_2012} photometric metallicity relations, respectively; they reported a final value of $[{\rm Fe/H}] = -0.01$\,dex.  \cite{Neves+Bonfils+Santos+etal_2014} slightly revised this value to $[{\rm Fe/H}] = -0.04 \pm 0.09$\,dex, again based on HARPS high signal-to-noise ratio, high-resolution optical spectroscopy. \cite{Nakajima+Tsuji+Takeda_2015} inferred approximately solar abundances of carbon and oxygen (depending on the adopted solar abundance pattern) and a slightly super-solar C/O ratio of $0.68\pm0.12$ from a high-resolution IRCS spectrum \citep{Nakajima+Tsuji+Takeda_2015}.  

There are also two older spectroscopically determined metallicities: \citet{Schiavon+Barbuy+Singh_1997} obtained $[{\rm Fe/H}] \approx -0.2$\,dex from the $\sim$1-$\mu$m FeH band, while \citet{Mould_1978} derived $[{\rm M/H}] \approx 0.15$\,dex from Fourier transform spectroscopy of aluminum, calcium, and magnesium lines.  None of the spectroscopic measurements support the very low metallicity of $-0.5$\,dex suggested by some of the early spectroscopic fits to either the host star or Gl~229B \citep{Leggett+Hauschildt+Allard+etal_2002}.

\subsection{Rotation and Activity} \label{subsec:rotation_activity}

Gl~229A shows chromospheric and coronal activity associated with its magnetic dynamo. The magnetic field heats regions above the stellar photosphere, causing a temperature inversion. Ca\,{\sc ii} HK lines originating in the hotter chromosphere appear narrowly in emission above the coolor photosphere, and serve as a proxy for magnetic activity \citep{Noyes+Hartmann+Baliunas+etal_1984}. Farther from the photosphere, the hot corona emits thermal X-rays to offer another indirect probe of magnetic activity.  The stellar dynamo is believed to be powered by rotation \citep[e.g.,][]{Parker_1955}. Stars slow their rotation as they age due to coupling with their magnetized winds, which rotate at constant angular frequency out to the Alfv\'en radius.  This spin-down provides an age indicator through gyrochonology \citep{Skumanich_1972,Barnes_2007}.

Chromospheric activity may be parametrized by $R^\prime_{\rm HK}$, an estimate of the ratio of the flux in the narrow HK emission lines to that in the underlying continuum \citep{Noyes+Hartmann+Baliunas+etal_1984}. Higher (less negative) $\log{R^\prime_{\rm HK}}$ values suggest youth. 
For Gl~229A, \cite{Astudillo-Defru+Delfosse+Bonfils+etal_2017} calibrated $\log R^\prime_{\rm HK} = -4.69$\,dex; the method of \cite{Brandt+Kuzuhara+McElwain+etal_2014} gives $\log R^\prime_{\rm HK} = -4.84$\,dex.  Adopting the higher activity index would push the age slightly lower, but well within the uncertainties in the conversion between activity and age \citep{Mamajek+Hillenbrand_2008}.  Gl~229A was also detected by {\it ROSAT} \citep{Voges+Aschenbach+Boller+etal_1999}.  The observed X-ray counts imply an activity index, the logarithmic ratio of X-ray to  bolometric flux, of $\log R_X = -5.14$\,dex.  There is no directly measured rotation period for Gl~229A; \cite{Astudillo-Defru+Delfosse+Bonfils+etal_2017} inferred a period of 25 days from measured chromospheric emission using correlations between rotation and activity. 

Gyrochronology for Gl~229A requires an extrapolation of the \cite{Mamajek+Hillenbrand_2008} relations to lower stellar masses, where $R^\prime_{\rm HK}$ and $R_X$ are poorly calibrated.  Still, Figures 9 and 10 of \cite{Mamajek+Hillenbrand_2008} suggest that gyrochronology remains viable down to early-M stars like Gl~229A.  We use the Bayesian age dating described in \cite{Brandt+Kuzuhara+McElwain+etal_2014} to obtain our activity-based age constraints; this method is based on the relations of \cite{Mamajek+Hillenbrand_2008}.  We first convert $R^\prime_{\rm HK}$ and $R_X$ to the Rossby number, and then to a rotation period using the convective overturn time as a function of color given in \cite{Noyes+Hartmann+Baliunas+etal_1984}.  The rotation period implied by $\log R_X = -5.14$\,dex is 30 days, while $\log R^\prime_{\rm HK} = -4.84$\,dex implies 47 days.  We then use the gyrochronological relations of \cite{Mamajek+Hillenbrand_2008} together with the weighted average of these rotation periods, 42 days, as described in \cite{Brandt+Kuzuhara+McElwain+etal_2014}.  

\cite{Angus+Morton+Foreman-Mackey+etal_2019} have re-calibrated gyrochronology down to early-M dwarfs, below the mass of Gl~229A.  Unfortunately, Gl~229A lacks a measured rotation period, and recalibrated coronal and chromospheric activity proxies for stellar rotation are not available. 
The gyrochronology relations in \cite{Mamajek+Hillenbrand_2008} give an age of 3.6~Gyr for $B-V=1.48$ and $P=42$~days, while those of \cite{Angus+Morton+Foreman-Mackey+etal_2019} give an age of 2.5~Gyr for {\it Gaia} $B_P - R_P = 2.08$\,mag and the same rotation period.  We propagate an uncertainty of 0.16 in Rossby number, as \cite{Mamajek+Hillenbrand_2008} suggest for Solar-type dwarfs with only $R^\prime_{\rm HK}$, to obtain $P_{\rm rot} = 42 \pm 5$~days.  Fixing the mass of Gl~229 to 0.54~$M_\odot$ as we determine in our dynamical fit, and fixing $B_P-R_P=2.08$\,mag, the gyrochronological relations of \cite{Angus+Morton+Foreman-Mackey+etal_2019} give an age of $2.6 \pm 0.5$~Gyr.  

\begin{figure}
    \centering
    \includegraphics[width=\linewidth]{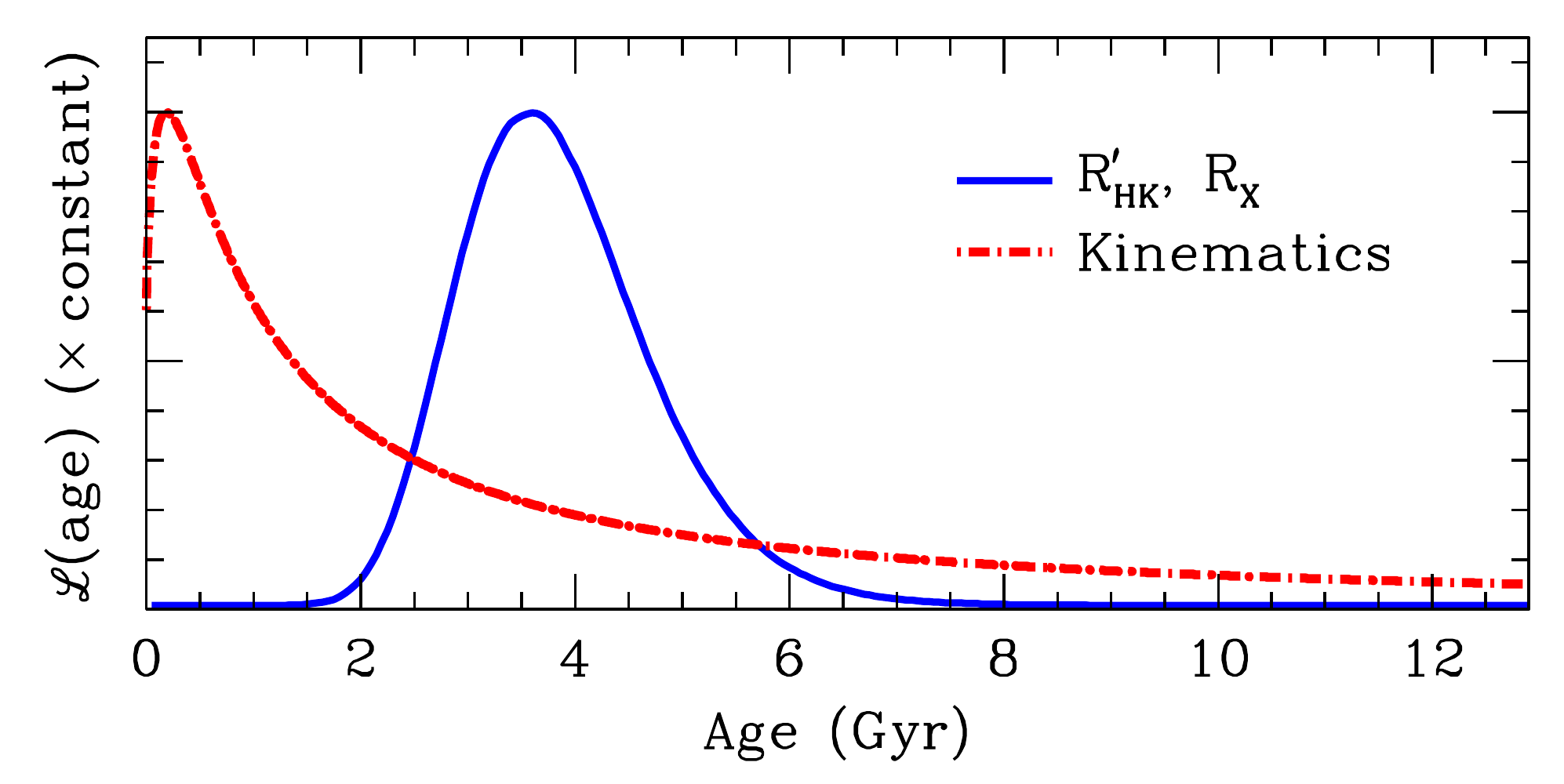}
    \caption{Age of Gl~229A from chromospheric and coronal activity (solid blue line) and from its  cold, thin disk kinematics (dot-dashed red line).  Section \ref{subsec:rotation_activity} describes the calculation of the activity age, while Section \ref{subsec:kinematics} describes the calculation of the kinematic age.  Both age indicators are consistent with an intermediate age of a few Gyr, and disfavor very old ages.  }
    \label{fig:age}
\end{figure}

Figure \ref{fig:age} shows our age posterior based on the \cite{Brandt+Kuzuhara+McElwain+etal_2014} methodology.  It is broad, ranging from $\sim$2~Gyr to $\sim$6~Gyr.  Adopting the inferred rotation rate and/or $R^\prime_{\rm HK}$ measurements of \cite{Astudillo-Defru+Delfosse+Bonfils+etal_2017} would exclude the older ages in our distribution.  The PARSEC stellar models \citep{Bressan+Marigo+Girardi+etal_2012} are consistent with Gl~229A lying on the main sequence, and do not provide meaningful age constraints.

\subsection{Kinematics} \label{subsec:kinematics}

Gl~229A is a member of the kinematic thin disk, with $UVW$ velocities of $(12, -12, -12)$~km\,s$^{-1}$, and a vertical height $Z \approx 2$~pc below the Solar position \citep{Gaia_Astrometry_2018,Soubiran+Jasniewicz+Chemin+etal_2013,Rodriguez_2016}.  The Sun itself is within $\sim$20--25~pc of the disk midplane; its exact position depends on the adopted definition of the midplane and the tracers used \citep{Bland-Hawthorn+Gerhard_2016,Karim+Mamajek_2017,Yao+Manchester+Wang_2017,Anderson+Wenger+Armentrout+etal_2019}.  Relative to the local standard of rest, Gl~229's $UVW$ velocities are approximately $(23, 0, -5)$~km\,s$^{-1}$ \citep{Schoenrich+Binney+Dehnen_2010}.  

The stellar disk increases in scale height and velocity dispersion as a function of age \citep[e.g.][]{Holmberg+Nordstrom+Andersen_2007}, likely due to interactions with molecular clouds and transient spiral structures \citep[][and references therein]{Binney+Tremaine_2008}.  We use the power-law fits of \cite{Holmberg+Nordstrom+Andersen_2007} to model the dynamical heating of the disk, taking the zero-age velocity dispersions to be 15, 10, and 5~km\,s$^{-1}$ in $U$, $V$, and $W$, respectively.  We add the power-law components in quadrature such that we match the $UVW$ dispersions of 50, 30, and 30~km\,s$^{-1}$ at 10~Gyr.  We take the velocity distribution near the disk midplane to be a multivariate Gaussian with these increasing dispersions.

With model velocity distributions as a function of age, we can estimate the likelihood of Gl~229's observed $UVW$ kinematics.  Our initial analysis neglects the increase in scale height, and corresponding decrease in midplane density, with population age.  Including this effect would make Gl~229's inferred kinematic age younger: dynamical heating makes the midplane overabundant in young stars relative to the thin disk as a whole.  We also adopt a uniform distribution of ages between 0 and 13~Gyr.  

The dot-dashed red line on Figure \ref{fig:age} shows our inferred kinematic age likelihood.  The 68\% and 95\% confidence upper limit on Gl~229's age are 4.2 and 10.8~Gyr, respectively, weak constraints that are consistent with the intermediate age favored by the star's activity.  Dividing the probability density by the population's $\sigma_W$ to account for the fact that density near the disk midplane decreases with time would reduce the upper limits to 2.0 and 8.3~Gyr.

The kinematic age constraint is weak; the peak at young ages is due to Gl~229's location near the midplane and velocity close to the local standard of rest. Taken together, our activity and kinematic analyses disfavor ages $\lesssim$2~Gyr and $\gtrsim$6~Gyr.  Multiplying the two probability distributions would yield a distribution closely resembling the activity age.  Gl~229A's low levels of chromospheric and coronal activity disfavor a very young system age much more strongly than kinematics favor it.  The two distributions are consistent with one another at a level equivalent to 1-$\sigma$: 33\% of the kinematic age distribution lies at ages above median activity age of 3.8~Gyr.

\section{Radial Velocities and Astrometry} \label{sec:rv_imaging}

\subsection{Radial Velocity Monitoring}

Gl~229 was observed using the HIRES \'echelle spectrograph \citep{Vogt+Allen+Bigelow+etal_1994} at Keck Observatory as part of the Lick-Carnegie Exoplanet Survey \citep[LCES,][]{Butler+Vogt+Laughlin+etal_2017}.  \cite{Butler+Vogt+Laughlin+etal_2017} published 47 radial velocity measurements taken between December 1996 and December 2013 with a median uncertainty of 1.33~m\,s$^{-1}$.  These spectra were reduced and calibrated using the same method and pipeline as the California Planet Survey \citep{Howard+Johnson+Marcy+etal_2010}.

The radial velocities for Gl~229 show a shallow linear trend of $0.3 \pm 0.1$~m\,s$^{-1}$\,yr$^{-1}$.  \cite{Tuomi+Jones+Barnes+etal_2014} reported a planet-mass companion at a radial velocity semi-amplitude of 4~m\,s$^{-1}$ and a period of $\sim$470 days using $\sim$7 years of HARPS and UVES spectra.  As we discuss in Section \ref{sec:orbitfit}, we see no evidence for this planet in our longer HIRES time series.

\subsection{Relative Astrometry from {\it HST} WFPC2} \label{subsec:hst_relast}

Gl~229B was observed between 1995 and 2000 using the planetary camera (PC) component of the Wide Field and Planetary Camera 2 (WFPC2) aboard the {\sl Hubble Space Telescope} ({\sl HST}).  \cite{Golimowski+Burrows+Kulkarni+etal_1998} presented astrometry from the first three epochs (1995--1996). The 1999--2000 measurements (from Proposal 8290, PI Christopher Burrows)  
have never been published but are available in the {\sl HST} archive.  We have (re-)derived all of the astrometry, as described below, and used the latest distortion correction provided in archival headers.

\begin{figure}
    \centering
    \includegraphics[width=\linewidth]{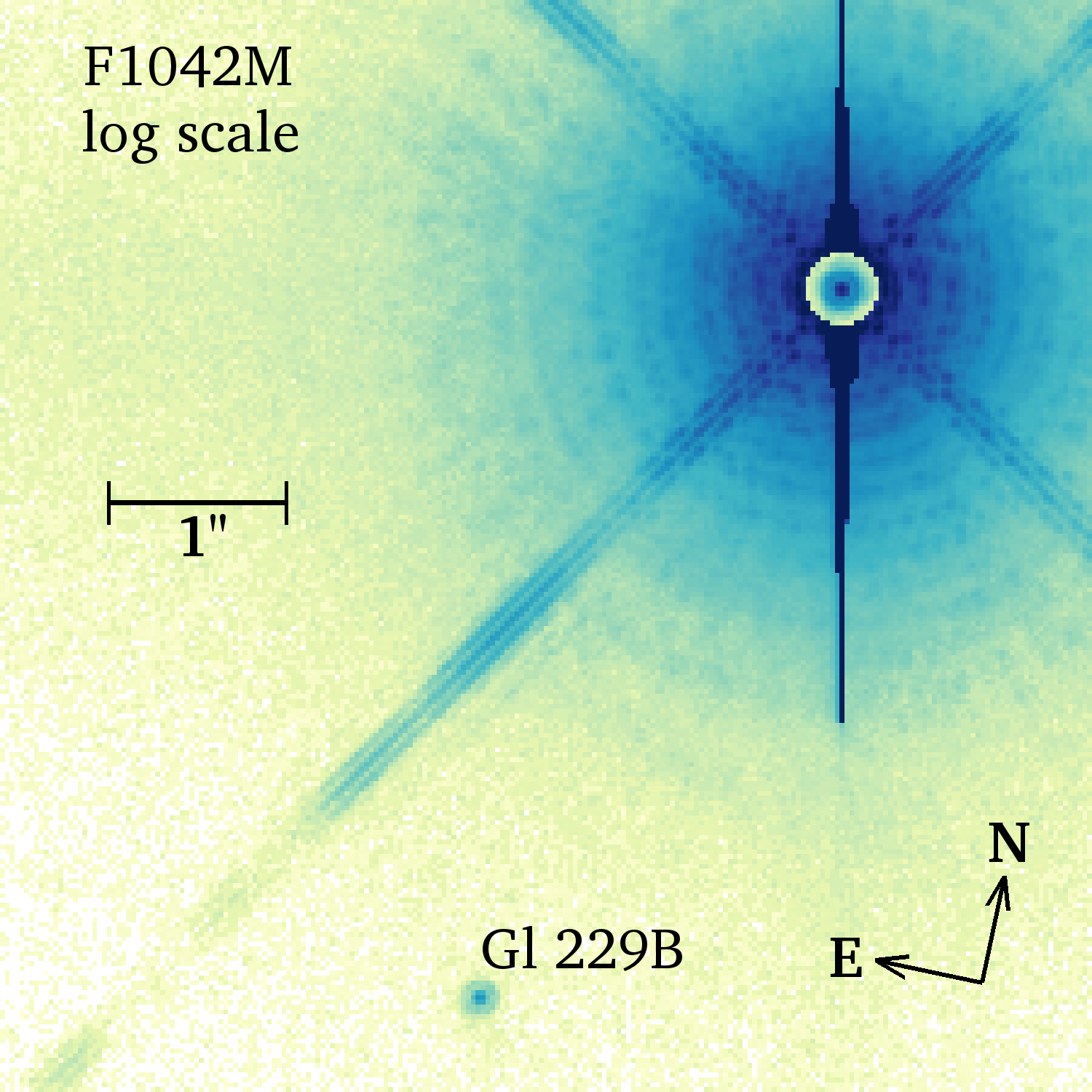}
    \caption{Image of Gl~229AB as seen in 1999 by WFPC2 on {\it HST} through the $\sim$1~$\mu$m filter F1042M. The observations consisted of short exposures (with Gl~229A unsaturated) and long exposures (with a good signal-to-noise ratio on Gl~229B).  The figure shows a composite of four dithered long exposures.  The inner, saturated region has been replaced by data from a short exposure. }
    \label{fig:Gl229_HST}
\end{figure}

{\sl HST} data were taken in a combination of short, 1.6-second exposures (to avoid saturating Gl~229A) and long, 400- to 500-second exposures to achieve a good signal-to-noise ratio on Gl~229B. Nearly all of these exposures were taken with the F1042M filter centered at 1.02~$\mu$m; a few images taken with the F835W filter produce consistent results. The frames were dithered by $\approx$5 pixels in both the horizontal and vertical directions.  
Figure \ref{fig:Gl229_HST} shows a composite {\it HST}/WFPC2 image from 1999 produced using four dither offsets; Gl~229B is visible in the lower-left.  We have replaced the saturated core of Gl~229A with data from a short, unsaturated image.
We measure relative astrometry using the point-spread function (PSF) of Gl~229A as our template to fit Gl~229B. 

We fit for the relative positions of Gl~229A and Gl~229B in pairs of exposures with one object unsaturated in each.  We use all such pairs from a given sequence of observations as individual measurements of the separation.  The best fit solution was determined by maximizing the cross-correlation between the interpolated PSF of Gl~229A and the PSF of Gl~229B, which is made possible by the relatively good sampling afforded by the WFPC2-PC pixels. We do not subtract the faint outer PSF halo of Gl~229A when fitting for the location of Gl~229B. We also checked reversing the object whose PSF we interpolated; the results do not change, implying that the slightly undersampled PSF is not a limiting factor in our analysis (at $\lambda = 1.02~\mu$m a PC pixel is $1.93 \lambda/D$). Subtracting the saturated PSF of Gl~229A rotated by 180$^\circ$ changes our derived astrometry by no more than 0.4~mas in separation and $0.\!\!^\circ$003 in position angle, both of which are much smaller than our uncertainties.  We then derive approximate offsets from the center of the frame (the origin of the distortion solution) and compute the difference of the distortion-corrected displacements from this center.  Finally, we add the dither offset recorded in the archival {\it HST} data.  In the absence of a distortion correction, this procedure is equivalent to converting the displacement in pixels to a displacement in RA and Dec using the linear transformation given in the file header.  Including the distortion correction changes the results by $\sim$1--10 mas depending on the location of Gl~229A in the field-of-view.  

We fit for the relative positions of Gl~229A and B in several pairs of exposures on each date; the scatter of these fitted positions gives one estimate of the error.  We can obtain another estimate by comparing offsets derived from dates closely spaced in time.  All of the results are consistent with the scatter being independent noise realizations in each pair of frames (and imply extremely reliable dithering offsets).  For each date, we report the astrometry averaged over all long/short exposure pairs, and adopt as our initial uncertainty estimate the standard deviation of offsets divided by $\sqrt{N-1}$, where $N$ is the number of exposure pairs on a given date.  Several sets of observations were taken just days apart.  In each such case, we report a single relative astrometry point.  The date-to-date scatter in the derived astrometry is $\lesssim$1~mas after dividing by $\sqrt{N-1}$.

Our error budget is dominated by factors other than photon noise or the scatter in our fitted centroid positions.  The distortion correction of WFPC2 is accurately known, but may have uncertainties, variations, and systematics at the $\sim$1~mas level \citep{Anderson+King_2003}.  We correct for the 34-th row artifact \citep{Anderson+King_1999}, which can also reach $\sim$1~mas. While {\sl HST} was dithered within a sequence of exposures, this dithering was by an approximately integer number of pixels, so we must also address a possible underestimation of the uncertainty due to the poor subpixel sampling of the PSFs (and hence, of the pixel response function).  We do this using a sample of 32 low-mass stars observed in the same filter and instrument configuration as Gl~229.  These 32 stars had -to-noise ratios ranging from values slightly lower than that of Gl~229B to values approaching that of Gl229A.  Our sample produces $32 \times 31$ pairs of images to which we can fit 31 relative offsets using the same method we apply to Gl~229AB.  The root-mean-square scatter of the individual offsets against these fitted offsets is 1~mas in both horizontal and vertical directions.  As one last consistency check, we computed the root-mean-square scatter in the positional difference $\Delta x_{ab} + \Delta x_{bc} - \Delta x_{ac}$; the result is consistent with $\sqrt{3}$ times the 1~mas root-mean-square scatters we found when comparing single offsets with average offsets.  We adopt 1.4~mas in separation uncertainty to account for the uncertain distortion correction and the effects of pixel sampling, combining 1~mas for each in quadrature.

We obtain our uncertainties in the position angle in a similar way as for separation.  We first divide the extra 1.4~mas in separation uncertainty by the measured separation to produce an uncertainty in position angle.  We then need to correct the orientation of the field.  For this purpose, we use the drizzled images from the {\sl HST} archive, which incorporate all known offsets between the chips, geometric distortions, and artifacts.  We then identify between five and seven stars in each field that are present in {\sl Gaia} DR2 \citep{Gaia_Astrometry_2018}.  Propagating the {\sl Gaia} positions back to the {\sl HST} epochs provides an astrometric reference field with typical positional uncertainties of $\sim$1~mas across the 2$.\!^\prime5$ WFPC2 field.  We do not apply a correction for parallactic motion.  All but one of the reference stars have parallaxes $\lesssim$1~mas, while the last one has a parallax of 4~mas.  We fit two-dimensional Gaussians to the stellar positions on the drizzled images and compute the best-fit rotation and translation between the coordinates of the stars in {\sl Gaia} DR2 and those measured on WFPC2.  After applying this rotation and translation, the remaining positional residuals between WFPC2 and {\it Gaia} are a few mas, which projects to an angle of $\sim$0$.\!\!^\circ$002.  We include an additional uncertainty of 0$.\!\!^\circ$01 to account for any variations in chip positions or orientations (though these are expected to be very small, \citealt{Gonzaga+Biretta+etal_2010}) and for possible systematic offsets resulting from spatial variations in the point-spread function.  

Four of the {\sl HST} epochs---1996~May, 1999~May, 2000~May, and 2000~November---used the same pair of guide stars.  The other two observations each used different pairs of guide stars, and might be misaligned with one another and/or with the other four epochs even after the drizzling correction.  Our position angle uncertainties could be underestimated for these two epochs, 1995 November and 1996 November.  The observations from 1995 November have an especially large angular offset of $0.\!\!^\circ17$ implied by the field stars in the drizzled images (the angular offsets are $\lesssim$0.$\!\!^\circ$03 for the other epochs).

The systematic differences between our relative astrometry and that of \cite{Golimowski+Burrows+Kulkarni+etal_1998} are $\sim$15~mas in separation and $\sim$0.\!\!$^\circ$3 in position angle.  The separation differences are $\sim$1.5$\sigma$-significant, but the position angle shows a $\sim$3$\sigma$ offset when we calibrate to {\sl Gaia} DR2. 
The $\chi^2$ of the projected separations in our best-fit orbit is 3.0 for seven measurements, suggesting that our $\sim$1.5~mas errors are conservative.  For position angles, the first three epochs used three different sets of guide stars.  The first epoch has a substantial $0.\!\!^\circ 17$ offset between the orientation given in the header and that computed from the {\it Gaia} reference stars.  Despite uncertainties a factor of $\sim$6 lower than those given by \cite{Golimowski+Burrows+Kulkarni+etal_1998}, we retain an acceptable $\chi^2 = 11.8$ in position angle for our best-fit orbit (with seven measurements).  Doubling the position angle uncertainties for 1995 November and 1996 November, when different guide stars were used, decreases the best-fit $\chi^2$ on position angle to 4.9 and has a negligible impact on our results.

\begin{deluxetable*}{lccccccr}
\tablewidth{0pt}
\tablecaption{Relative Astrometry}
\tablehead{
    \colhead{Mean Date} &
    \colhead{$\rho$} &
    \colhead{$\sigma_\rho$} &
    \colhead{PA} &
    \colhead{$\sigma_{\rm PA}$} &
    \colhead{Instrument} &
    \colhead{Number of} &
    \colhead{Number of}
 \\
    \colhead{yyyy-mm-dd} &
    \multicolumn{2}{c}{(mas)} &
    \multicolumn{2}{c}{(degrees)} &
    \colhead{} &
    \colhead{Epochs\tablenotemark{*}} &
    \colhead{Image Pairs}
}
\startdata
1995-11-17 & 7777.0 & 1.7 & 163.224 & 0.015 & {\it HST} WFPC2\tablenotemark{$\dagger$} & 1 & 8 \\
1996-05-25 & 7732.0 & 2.0 & 163.456 & 0.019 & {\it HST} WFPC2 & 1 & 4 \\
1996-11-09 & 7687.7 & 1.5 & 163.595 & 0.015 & {\it HST} WFPC2\tablenotemark{$\dagger$} & 1 & 8 \\
1999-05-26 & 7458.3 & 1.6 & 164.796 & 0.015 & {\it HST} WFPC2 & 2 & 32 \\
2000-05-26 & 7362.8 & 1.6 & 165.244 & 0.016 & {\it HST} WFPC2 & 3 & 48 \\
2000-11-16 & 7316.9 & 1.6 & 165.469 & 0.016 & {\it HST} WFPC2 & 3 & 48 \\
2011-03-26 & 6210  & 10  & 171.2  & 0.1  & HiCIAO & 1 & 3
\enddata
\tablenotetext{*}{Observations within $\pm$2 days of the mean date are combined into a single measurement. }
\tablenotetext{$\dagger$}{Data from 1995 November and 1996 November used different guide stars from one another and from the other four {\it HST} epochs.}
\label{tab:companionastrometry}
\end{deluxetable*}

The first six lines of Table \ref{tab:companionastrometry} list the relative astrometry we derive for Gl~229B from archival {\it HST} WFPC2 images.  These data show clear evidence of orbital motion.

\subsection{Relative Astrometry from HiCIAO}

After intensive monitoring by {\it HST} between 1995 and 2000, Gl~229 has little published astrometry.  \cite{Geissler+Chauvin+Sterzik+etal_2008} did obtain 8.6-$\mu$m images in 2006 using VLT/NACO, but the astrometric precision of these measurements is poor ($\sigma_\rho = 50$~mas, $\sigma_{\rm PA} = 0.\!\!^\circ{}9$).  Gl~229 was also briefly observed as part of the SEEDS survey \citep{Tamura_2009} on 26 March 2011, using the HiCIAO instrument \citep{Suzuki+Kudo+Hashimoto+etal_2010} behind the adaptive optics system AO188 \citep{Hayano+Takami+Guyon+etal_2008,Minowa+Hayano+Watanabe+etal_2010} on the Subaru telescope.  These data were taken both in direct imaging mode (with a single broadband $H$ filter) and in polarized differential imaging mode, with the field split in half by a Wollaston prism in the fore-optics.  We use only the broadband direct images for our analysis, which we downloaded from the public Subaru archives.  The image rotator was turned on throughout this sequence; we do not need to correct for parallactic angle.

The HiCIAO data set comprises four short (1.5-second) exposures with a 1\% neutral density filter in place to avoid saturating Gl~229A, along with three saturated 60-second exposures.  We remove correlated read noise (``striping''), flat-field, and perform a distortion correction using ACORNS-ADI \citep{Brandt+McElwain+Turner+etal_2013}.  The data used to calibrate the distortion solution for March 2011 were of the globular cluster M5, and were taken during the same observing run as our images of Gl~229.  The HiCIAO astrometry is calibrated to distortion-corrected {\it HST} images taken with WFPC2, the same instrument we use for the relative astrometry of Gl~229 in Section \ref{subsec:hst_relast}.  The resulting third-order distortion correction has an internal fractional uncertainty of $10^{-4}$ in the plate scale (linear terms), or less than 1~mas at the $\sim$6$^{\prime\prime}$ separation of Gl~229B.  Uncertainties in the nonlinear terms of the distortion solution contribute $\lesssim$2~mas of uncertainty to the separation given the observed positions of Gl~229A and Gl~229B on the HiCIAO detector.

We separately centroid Gl~229A in the unsaturated frames and Gl~229B in the saturated frames, assuming their positions on the detector to remain fixed when changing the exposure time and removing the neutral density filter.  \cite{Kuzuhara+Tamura+Kudo+etal_2013} verified this approach using images of a pinhole mask at the focal plane of HiCIAO,
with and without ND filters; they found the systematic offsets to be $\lesssim$3~mas.  \cite{Kuzuhara+Tamura+Kudo+etal_2013}  further confirmed the small offset using three binaries in the SEEDS sample.  The approach of interspersing saturated and unsaturated images was used for many targets in the latter part of the SEEDS survey \citep{Uyama+Hashimoto+Kuzuhara_etal_2017}.  In each case, we cross-correlate our images against the average HiCIAO PSFs distributed with ACORNS-ADI.  The formal uncertainties of this procedure are much less than a pixel.  Unfortunately, we do not have any background stars to validate the astrometry.
In sequences of HiCIAO data, frame-to-frame scatter was less than 1 pixel \citep{Brandt+McElwain+Turner+etal_2013}.  Visually, the center of the saturated star is consistent with its position in unsaturated images.  We adopt 1 pixel (or 10~mas) as our uncertainty in the separation of the star and its companion.  A 10~mas uncertainty projects to our adopted uncertainty of 0.$\!\!^\circ$1 in position angle.  The precision of the distortion corrected value of true north is $\sim$0$.\!\!^\circ$005; its value was stable to $\sim$0$.\!\!^\circ$03 during the SEEDS survey \citep{Brandt+McElwain+Turner+etal_2013}.

The last line of Table \ref{tab:companionastrometry} lists our HiCIAO measurement from March of 2011.  Despite its lower precision, the long time baseline between the HiCIAO observation and the {\it HST} images makes it a valuable constraint on the companion's orbit.  It does not drive our main result, however.  Quintupling the uncertainty on the HiCIAO astrometry reduces our best-fit mass in Section \ref{sec:orbitfit} by $\sim$0.5~$M_{\rm Jup}$, or $0.1\sigma$, while omitting the measurement entirely reduces the best-fit mass by $\sim$1~$M_{\rm Jup}$, or $0.2\sigma$.  We omit the 2006 VLT/NACO astrometry from our analysis, as its precision is a factor of 5--10 worse than that of HiCIAO and it offers no additional time baseline.

\subsection{Absolute Astrometry from the HGCA} \label{subsec:absast}

We take our absolute stellar astrometry from the HGCA.  This catalog gives nearly instantaneous proper motions at the {\it Hipparcos} and {\it Gaia}~DR2 epochs, along with the difference in positions between the two catalogs divided by the time baseline.  Each measurement also has a central epoch similar to, but slightly distinct from, the catalog epochs of 1991.25 and 2015.5.  

\begin{deluxetable*}{lccccccr}
\tablewidth{0pt}
\tablecaption{Absolute Stellar Astrometry}
\tablehead{
    \colhead{Mission} &
    \colhead{$\mu_{\alpha*}$} &
    \colhead{$\sigma[\mu_{\alpha*}]$} &
    \colhead{$\mu_{\delta}$} &
    \colhead{$\sigma[\mu_\delta]$} &
    \colhead{Corr$[\mu_{\alpha*},\mu_\delta]$} &
    \colhead{$t_{\alpha*}$} &
    \colhead{$t_\delta$} \\
    \colhead{} &
    \multicolumn{2}{c}{(mas\,yr$^{-1}$)} & 
    \multicolumn{2}{c}{(mas\,yr$^{-1}$)} &
    \colhead{} &
    \multicolumn{2}{c}{(year)}
    }
\startdata
\Hipparcos        & $-137.34$   & 0.58                  & $-713.85$   & 0.83                  & $-0.33$ & 1991.05 & 1991.29 \\
\Hipparcos--\Gaia~DR2 & $-136.514$ & 0.020                & $-714.967$ & 0.031                & $-0.18$ & \nodata & \nodata \\
\Gaia~DR2             & $-135.99$   & $0.19$ & $-719.09$   & 0.27 & $-0.25$ & 2015.04 & 2015.26
\enddata
\label{tab:hip_gaia}
\end{deluxetable*}

Table \ref{tab:hip_gaia} lists the catalog astrometry for Gl~229A.  The HGCA performs a full cross-calibration of \Hipparcos{} and \Gaia~DR2 \citep{Gaia_General_2016,Gaia_General_2018,Gaia_Astrometry_2018} including locally variable frame rotation, an optimized linear combination of the two \Hipparcos{} reductions \citep{ESA_1997, vanLeeuwen_2007}, and error inflation.  We refer the reader to \cite{Brandt_2018} for a detailed discussion of the catalog.  

Gl~229A appears to be an excellent fit in both {\it Hipparcos} and {\it Gaia} DR2. It has no rejected data in either {\it Hipparcos} reduction, and has slightly negative goodness-of-fit metrics (i.e.~better than expected on average for Gaussian errors) in both the \cite{ESA_1997} and \cite{vanLeeuwen_2007} catalogs.  Gl~229A likewise has no rejected astrometric data in {\it Gaia} DR2 and has a renormalized unit weight error of 0.98.

Gl~229B is not present in {\it Gaia} DR2; its $I$-band brightness of $\sim$19.5 \citep{Golimowski+Burrows+Kulkarni+etal_1998} and proximity to the bright star Gl~229A may have placed it just beyond {\it Gaia}'s reach.  The T dwarf companion might be present in future \Gaia{} data releases; its position relative to Gl~229A (and potentially even its relative proper motion) would further constrain the system's orbit.

\section{Orbital Fit} \label{sec:orbitfit}

\subsection{Single-Epoch Approximation}

As a first step, we use the method described in Section 5 of \cite{Brandt+Dupuy+Bowler_2018} to compute Gl~229B's dynamical mass.  This method fits the relative astrometry and radial velocity using quadratics in space and time, respectively.  It then computes the instantaneous projected separation and radial velocity acceleration at a characteristic epoch for the absolute astrometry of the HGCA.  We then calculate the companion mass using
\begin{equation}
    M = \frac{\rho^2 \left(a_{\alpha\delta}^2 + a_{\rm RV}^2\right)^{3/2}}{\varpi^2 G a_{\alpha\delta}^2},
\end{equation}
where $\rho$ is the projected separation, $\varpi$ is the parallax, $a_{\alpha\delta}$ and $a_{\rm RV}$ are the astrometric and radial velocity accelerations, respectively, and $G$ is the gravitational constant.

\cite{Tuomi+Jones+Barnes+etal_2014} reported a periodic radial velocity signal with a semi-amplitude of $\sim$4~m\,s$^{-1}$, a period of $\sim$470~days, and near-zero eccentricity.  However, fitting a linear trend plus a zero-eccentricity planet to our HIRES radial velocities does not improve the fit significantly over a linear trend alone.  Fitting for a zero-eccentricity planet of 4~m\,s$^{-1}$ semi-amplitude greatly degrades the quality of the fit, increasing the required ``jitter'' from about 3~m\,s$^{-1}$ to nearly 4~m\,s$^{-1}$ even after optimizing over the planet's phase.  \cite{Butler+Vogt+Laughlin+etal_2017} did not see any periodic signal in their own analysis of the HIRES data.

Without independent confirmation of the planet reported by \cite{Tuomi+Jones+Barnes+etal_2014}, and with a significant degradation of our residuals at the best-fit orbital parameters reported by \cite{Tuomi+Jones+Barnes+etal_2014}, we exclude this companion from our analysis.  The 17-year baseline of our data and its uneven sampling limit the effect to excluding a real planet with these properties to $\lesssim$0.5~m\,s$^{-1}$\,yr$^{-1}$.

\begin{deluxetable}{lcccr}
\tablewidth{0pt}
\tablecaption{Single-Epoch Approximate Mass of Gl~229B}
\tablehead{
    \colhead{Epoch} &
    \colhead{$\rho$} &
    \colhead{$a_{\alpha\delta}$} &
    \colhead{$a_{\rm RV}$} & 
    \colhead{$M_{\rm B}$} \\
    \colhead{} &
    \colhead{(mas)} &
    \multicolumn{2}{c}{(m\,s$^{-1}$\,yr$^{-1}$)} &    \colhead{($M_{\rm Jup}$)} 
}  
\startdata
2009.54 & $6404 \pm 14$ & $9.35 \pm 0.62$ & $0.07 \pm 0.55$ & $72 \pm 5$
\enddata
\label{tab:approx_mass}
\end{deluxetable}

Table \ref{tab:approx_mass} shows the results of the single-epoch approximation to the mass of Gl~229B.  We obtain a value of $72\pm5$~$M_{\rm Jup}$ that is much higher than previously estimated mass ranges, but marginally consistent with the maximum mass derived from theoretical considerations assuming an age of $\sim$10~Gyr \citep[e.g.,][]{Allard1996,Marley1996}.  The Gl~229AB system traced out a small fraction of an orbital arc in 15 years of astrometric monitoring; the assumptions required to compute a single-epoch mass are well-satisfied.  
We perform a self-consistent orbital analysis to verify the single-epoch result, obtain full posterior probability distributions on orbital parameters, and to check for goodness-of-fit.  

\subsection{A Full Orbital Fit}

We fit the orbit of Gl~229AB using the method described in \cite{Brandt+Dupuy+Bowler_2018}, with a few significant modifications.  As previously, we perform a parallel-tempering Markov Chain Monte Carlo (MCMC) analysis using \texttt{emcee} \citep{2005PCCP....7.3910E,Foreman-Mackey+Hogg+Lang+etal_2013}; our likelihood is calculated by comparing the measured separations, position angles, absolute astrometry, and radial velocities to those of a synthetic orbit and assuming Gaussian errors.  This method includes several nuisance parameters, 
including the precise proper motion and radial velocity of the system barycenter and the radial velocity jitter.  To reduce the number of parameters fit by \texttt{emcee}, we explicitly marginalize out the proper motion of the barycenter, the systemic radial velocity, and the parallax (adopting the {\it Gaia} DR2 value of $173.6955 \pm 0.0457$~mas as our prior).  In total, our analysis uses nine free parameters: the masses of the host star ($M_{\star}$) and companion ($M_{\rm comp}$), RV jitter, semimajor axis ($a$), inclination ($i$), PA of the ascending node ($\Omega$), mean longitude at a reference epoch ($t_{\rm ref}$) of 2455197.5~JD ($\lambda_{\rm ref}$; 2010 Jan 1 00:00 UT), and finally the eccentricity ($e$) and the argument of periastron ($\omega$) fitted as $\sqrt{e}\sin{\omega}$ and $\sqrt{e}\cos{\omega}$.

Our most important modification to the method of \cite{Brandt+Dupuy+Bowler_2018} is our use of epoch astrometry from both {\it Hipparcos} and {\it Gaia}, as described in Brandt et al.~(in preparation).  The epochs and scan angles of both {\it Hipparcos} and {\it Gaia} are publicly available. For the original {\it Hipparcos} reduction \citep{ESA_1997}, they are contained in the intermediate astrometric data hosted by the European Space Agency\footnote{\url{https://www.cosmos.esa.int/web/hipparcos/intermediate-data}}.  The intermediate data of the \cite{vanLeeuwen_2007} re-reduction were distributed on compact disc and are more difficult to obtain, but are available on request.  Predicted {\it Gaia} observations may be obtained from the Observation Forecast Tool\footnote{\url{https://gaia.esac.esa.int/gost/}}.
 
The scan angles and uncertainties provided with the intermediate {\it Hipparcos} astrometric data are sufficient to construct the covariance matrices used to fit for proper motion.  For {\it Gaia}, we assume that the along-scan uncertainties are the same for all observations.  We can then sample our synthetic orbits at the observed epochs and fit linear motion exactly as the processing teams did, using either the real covariance matrices or a close approximation.  We then compare the resulting positions and proper motions with the values given in \cite{Brandt_2018}.  This approach allows us to accurately reproduce the {\it Hipparcos} and {\it Gaia} measurements, even though {\it Gaia} epoch astrometry is currently unavailable and will not be published for several years.

\begin{figure}
\centerline{
\includegraphics[width=0.84\linewidth]{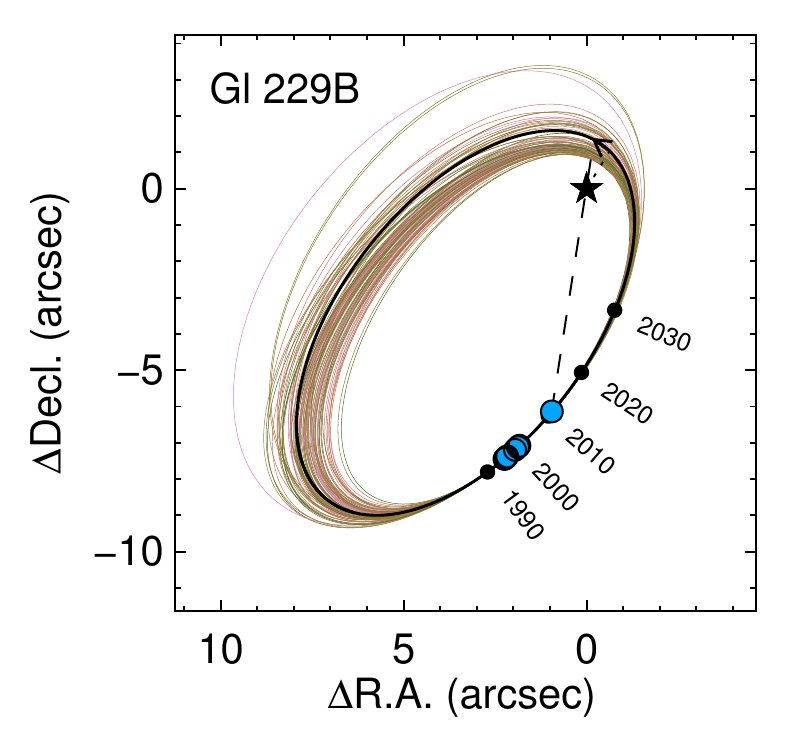}
\hskip -2.35 truein
\includegraphics[width=0.83\linewidth]{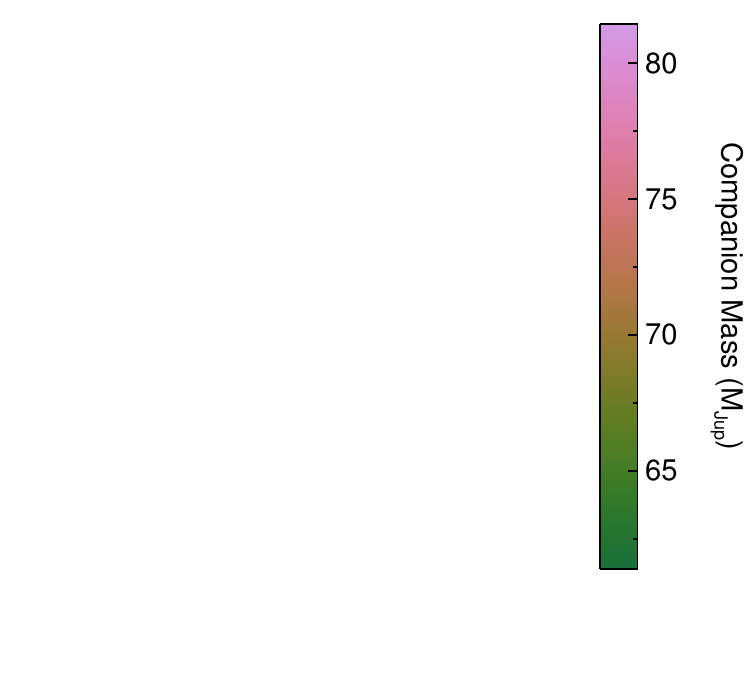}
}
\vskip 0.0 truein
\caption{Projected orbit of Gl~229B on the sky, with data used in the orbit fit shown as blue circles. The thick black line shows the orbit with the highest posterior probability density and thin lines are 100 orbits drawn randomly from our posterior distribution. They are colored according to companion mass from green (low mass) to pink (high mass).}
\label{fig:GL229-rd}
\end{figure}

\begin{figure*}
\centerline{
\includegraphics[width=0.3\linewidth]{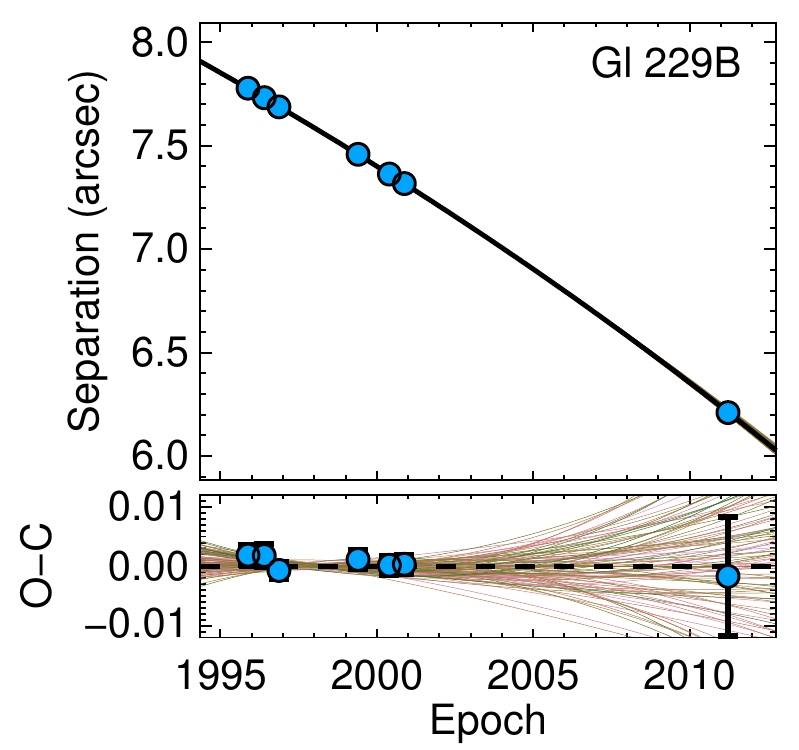}
\includegraphics[width=0.3\linewidth]{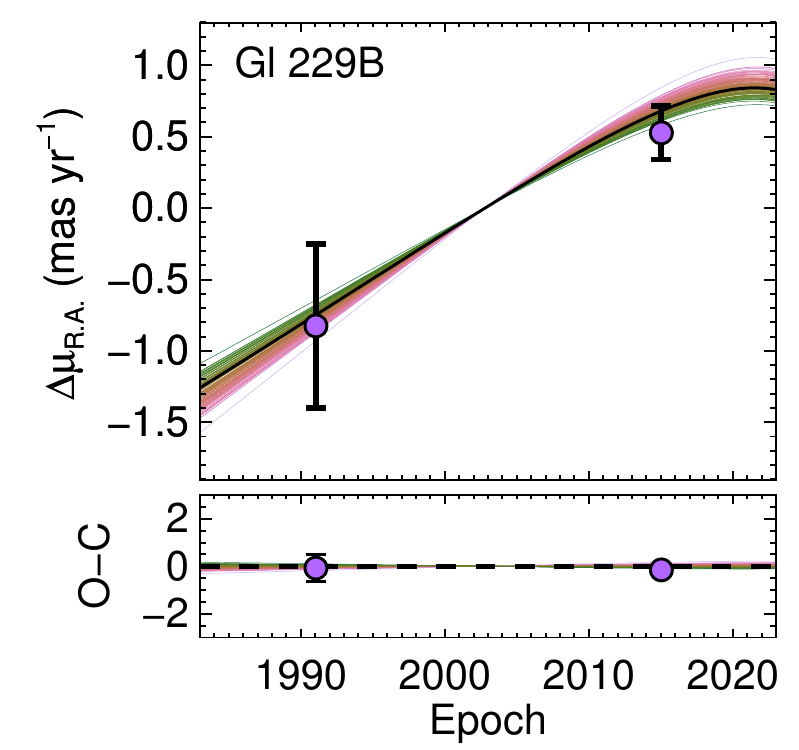}
\includegraphics[width=0.3\linewidth]{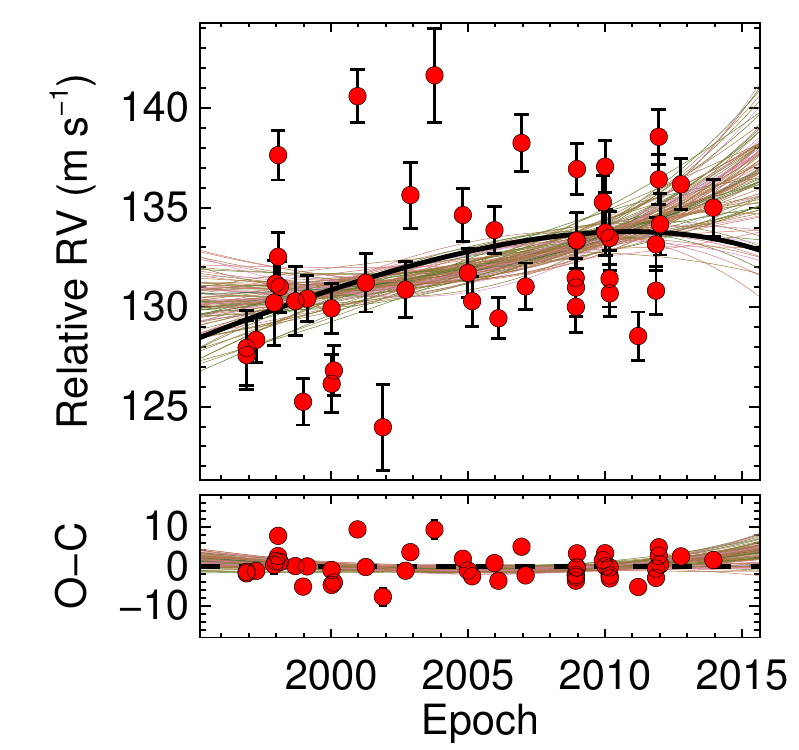}
}
\centerline{
\includegraphics[width=0.3\linewidth]{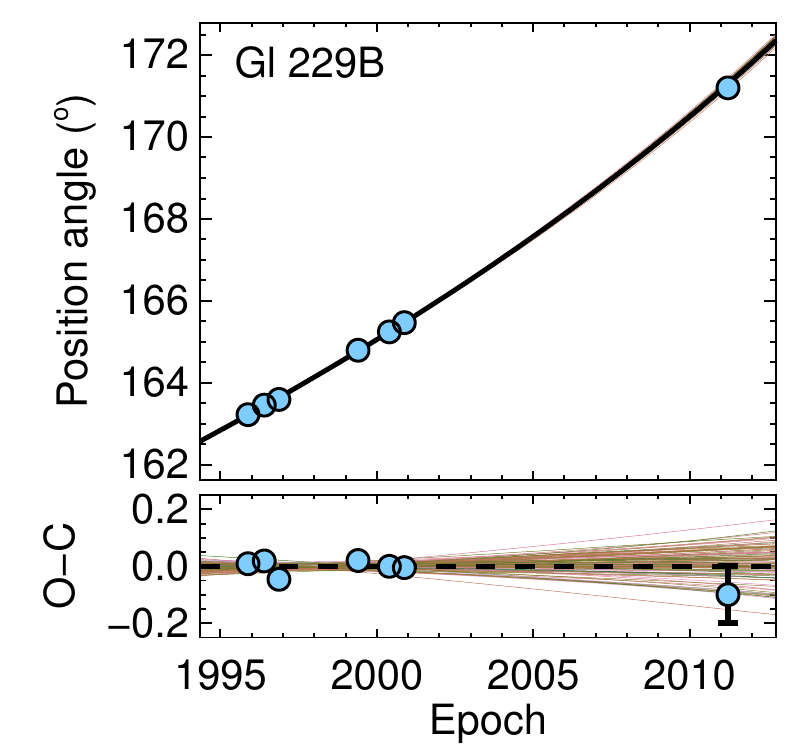}
\includegraphics[width=0.3\linewidth]{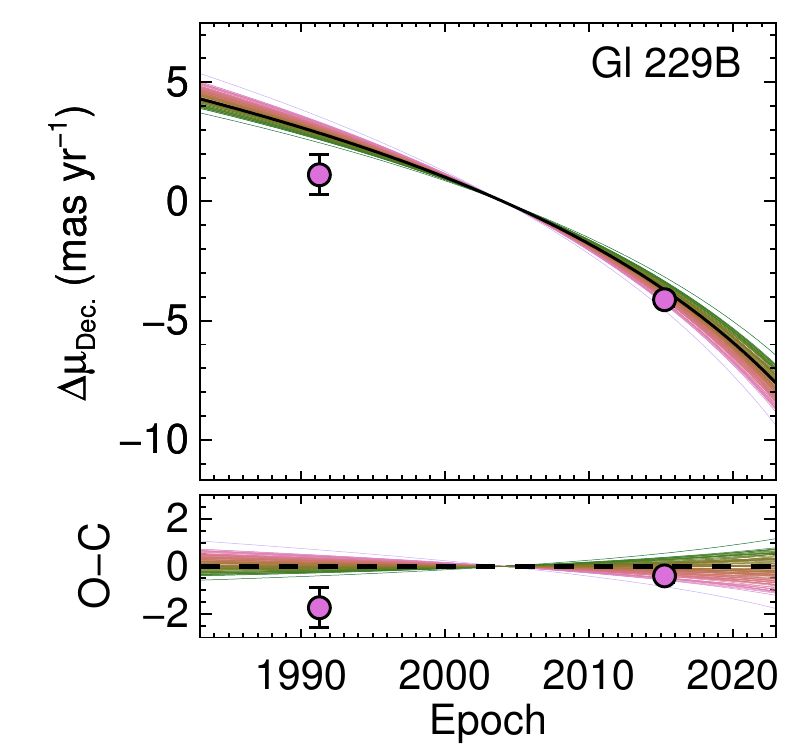}
\includegraphics[width=0.3\linewidth]{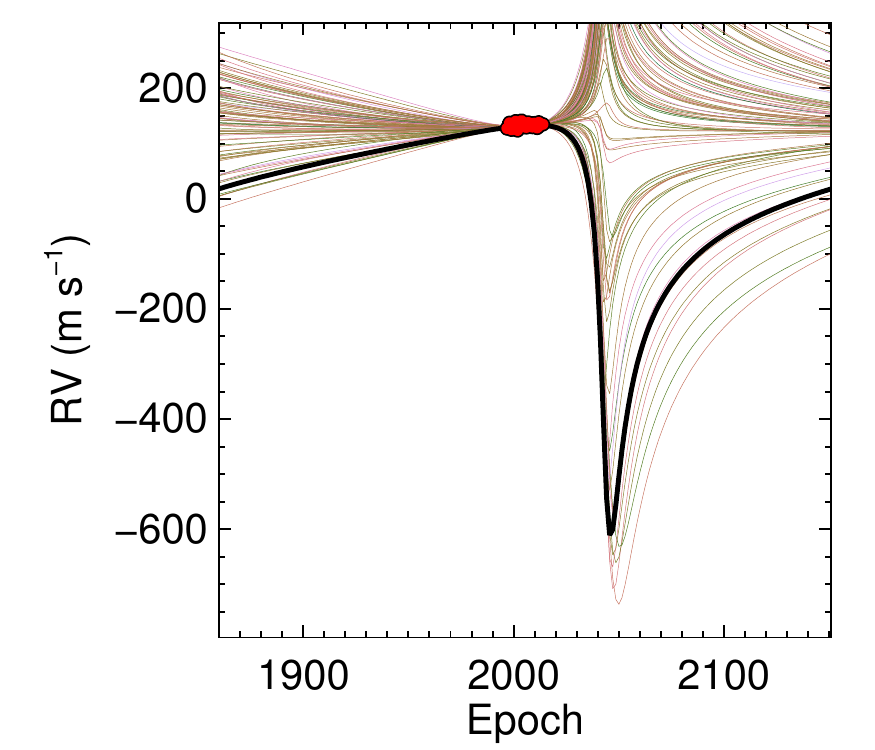}
}
\vskip 0.0 truein
\caption{Astrometry and RVs as a function of time for the Gl~229 system. The line colors and thicknesses have the same meaning as Figure~\ref{fig:GL229-rd}. Green lines correspond to lower companion masses, and pink lines are higher masses. Left: astrometry of Gl~229B relative to its host star from direct imaging. Middle: astrometric acceleration induced by Gl~229B on its host star, where $\Delta\mu$ is the difference between the \Hipparcos\ and \Gaia\ proper motions compared to the scaled positional difference between the two catalogs (i.e., the 25-year proper motion). Right: RVs of the host star measured from Keck (top) and the full RV orbit over three centuries.}
\label{fig:GL229-ast}
\end{figure*}

\begin{figure*}
\includegraphics[width=1.0\linewidth]{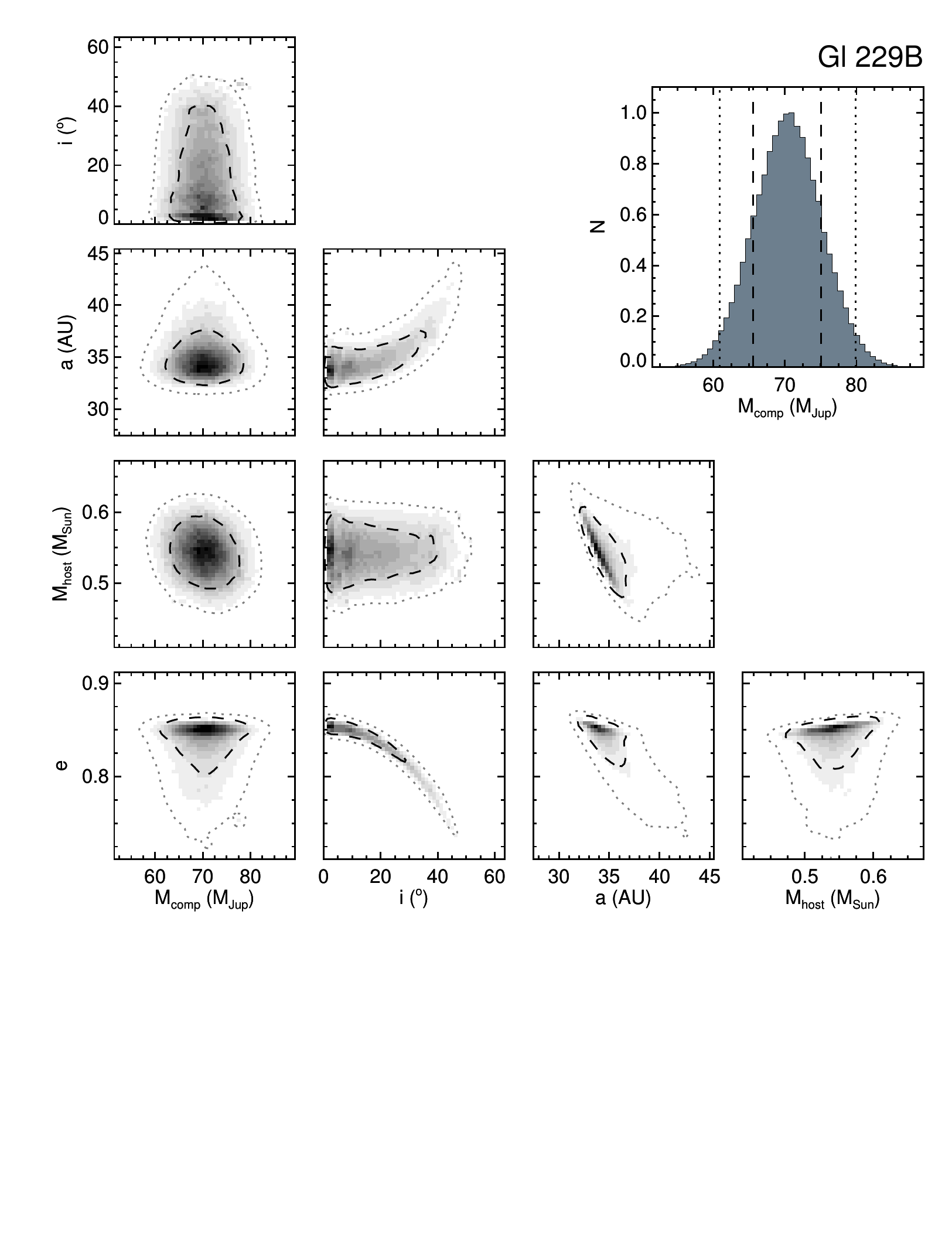}
\vskip -2.25 truein
\caption{Joint posterior distributions for selected orbital parameters. Dark dashed contours indicate 1$\sigma$ ranges and lighter dotted contours indicate 2$\sigma$ ranges. The top right panel shows the posterior distribution of the companion's mass.}
\label{fig:posterior1}
\end{figure*}

\begin{figure*}
]\includegraphics[width=1.0\linewidth]{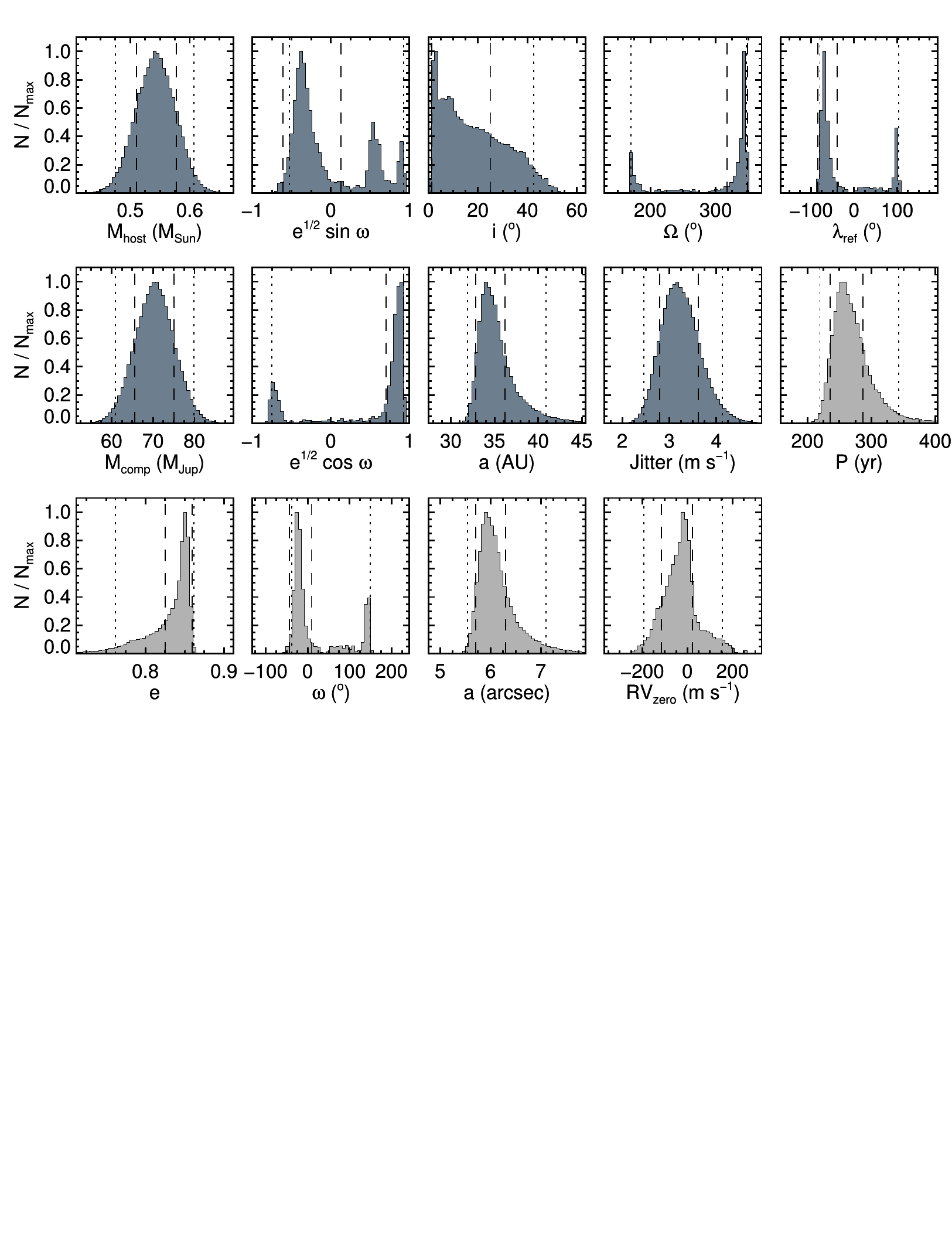}
\vskip -3.9 truein
\caption{Marginalized posterior distributions for fitted orbital parameters (dark gray histograms) and select parameters computed from these (light gray histograms).}
\label{fig:posterior2}
\end{figure*}

Figures \ref{fig:GL229-rd} and \ref{fig:GL229-ast}  show our orbital fits, using black lines to show the best-fit orbit (the one with the maximum posterior probability density).  Colored lines show 100 orbits randomly drawn from the MCMC posterior probability distributions, color-coded by Gl~229B's mass.  Figures~\ref{fig:posterior1} and \ref{fig:posterior2} show the posterior probability distributions and covariances of fitted parameters, as well as some parameters that can be computed from the directly fitted ones. The highest correlation is between eccentricity and inclination, where more face-on orbits have higher $e$ ($\approx$0.85), and neither of these parameters are strongly correlated with the mass of Gl~229B. For clarity, the nuisance parameters of RV jitter and the viewing angles $\omega$, $\Omega$, and $\lambda_{\rm ref}$ are not shown in Figure~\ref{fig:posterior1}; they also show no significant correlation with $M_{\rm comp}$. Table \ref{tab:posteriors} lists the median, 1$\sigma$, and 2$\sigma$ posterior intervals on all parameters.  Table \ref{tab:predictions} lists our predictions of the separation, position angle, and host star radial velocity from 1995 to 2030.

Our best-fit orbit has a $\chi^2$ of 3.0 for seven separation measurements, and $\chi^2=11.8$ for seven position angles.  The $\chi^2$ for position angles is dominated by the 1995 November and 1996 November epochs, which used different sets of guide stars than the other {\it HST} astrometric measurements.  Doubling the error bars on these two position measurements to $0.\!\!^\circ03$ reduces the $\chi^2$ on position angle to 4.9, and has a negligible effect on the posterior distributions of any of the fitted parameters.  

Our maximum posterior probability orbit has a $\chi^2$ of 7.4 for six absolute astrometry measurements.  We marginalize out two parameters for the proper motion of the system barycenter, reducing the number of degrees of freedom to four.  Our use of the proper motion to constrain the orbit further reduces the number of degrees of freedom, making a value of 7.4 somewhat higher than we would expect.  The probability of obtaining $\chi^2 \geq 7.4$ is 12\% assuming four degrees of freedom, and 6\% assuming three degrees of freedom.  Future {\it Gaia} data releases will enable additional consistency checks on the absolute astrometry.  However, as discussed in Section \ref{subsec:absast}, the astrometry of Gl~229A passes every internal consistency check in both {\it Hipparcos} and {\it Gaia}.

%%%%%%%%%%%%%%%%%%%%%%%%%%%%%%%%%%%%%%%%%%%%%%%%%%%%%%%%%%%

\begin{deluxetable*}{lccc}
\tablecaption{PT-MCMC Orbital Posteriors for Gl~229B \label{tbl:mcmc-GL229}}
\setlength{\tabcolsep}{0.10in}
\tabletypesize{\tiny}
\tablewidth{0pt}
\tablehead{
\colhead{Property}              &
\colhead{Median $\pm$1$\sigma$} &
\colhead{95.4\% c.i.}           &
\colhead{Prior}                 }
\startdata
\multicolumn{4}{c}{Fitted parameters} \\[1pt]
\cline{1-4}
\multicolumn{4}{c}{} \\[-5pt]
Companion mass $M_{\rm comp}$ (\Mjup)                                       & $70.4 \pm 4.8$                   &           61, 80           & $1/M$ (log-flat)                                                   \\[3pt]
Host-star mass $M_{\rm host}$ (\Msun)                                       & $0.54_{-0.03}^{+0.04}$           &         0.48, 0.61         & $1/M$ (log-flat)                                                   \\[3pt]
Semimajor axis $a$ (AU)                                                     & $34.7_{-1.9}^{+1.3}$             &         31.7, 40.1         & $1/a$ (log-flat)                                                   \\[3pt]
Inclination $i$ (\degree)                                                   & $13_{-12}^{+10}$                 &            1, 41           & $\sin(i)$, $0\degree < i < 180\degree$                             \\[3pt]
$\sqrt{e}\sin{\omega}$                                                      & $-0.21_{-0.29}^{+0.75}$          &      $-$0.54, 0.93         & uniform                                                            \\[3pt]
$\sqrt{e}\cos{\omega}$                                                      & $0.80_{-0.22}^{+0.13}$           &      $-$0.75, 0.93         & uniform                                                            \\[3pt]
Mean longitude at $t_{\rm ref}=2455197.5$~JD, $\lambda_{\rm ref}$ (\degree) & $-60_{-26}^{+49}$                &        $-$80, 103          & uniform                                                            \\[3pt]
PA of the ascending node $\Omega$ (\degree)                                 & $335_{-48}^{+13}$                &          171, 347          & uniform                                                            \\[3pt]
RV jitter $\sigma_{\rm jit}$ (m\,s$^{-1}$)                                  & $3.2\pm0.4$                      &          2.4, 4.1          & $1/\sigma_{\rm jit}$ (log-flat)                                    \\[3pt]
\cline{1-4}
\multicolumn{4}{c}{} \\[-5pt]
\multicolumn{4}{c}{Computed properties} \\[1pt]
\cline{1-4}
\multicolumn{4}{c}{} \\[-5pt]
Orbital period $P$ (yr)                                                     & $263_{-29}^{+21}$                &          217, 336          & \nodata                                                            \\[3pt]
Semimajor axis (mas)                                                        & $6030_{-330}^{+220}$             &         5510, 6970         & \nodata                                                            \\[3pt]
Eccentricity $e$                                                            & $0.846_{-0.015}^{+0.014}$        &        0.764, 0.864        & \nodata                                                            \\[3pt]
Argument of periastron $\omega$ (\degree)                                   & $-10_{-40}^{+50}$                &        $-$40, 150          & \nodata                                                            \\[3pt]
Time of periastron $T_0=t_{\rm ref}-P\frac{\lambda-\omega}{360\degree}$ (JD)& $2467400_{-500}^{+400}$          &      2466500, 2468900      & \nodata                                                            \\[3pt]
Mass ratio $q = M_{\rm comp}/M_{\rm host}$                                  & $0.123_{-0.013}^{+0.012}$        &        0.101, 0.150        & \nodata                                                            
\enddata
\tablecomments{The $\chi^2$ of relative astrometry is 3.0 for separations and 11.8 for PAs, with 7 measurements for each. The $\chi^2$ of the Hipparcos and Gaia proper motion differences is 7.4 for four measurements (after marginalizing out the barycenter's proper motion).}
\label{tab:posteriors}
\end{deluxetable*}
%%%%%%%%%%%%%%%%%%%%%%%%%%%%%%%%%%%%%%%%%%%%%%%%%%%%%%%%%%%%%%%%%%%%%%%

\begin{deluxetable}{lccr}
\tablewidth{0pt}
\tablecaption{Predicted Positions and Radial Velocities}
\tablehead{
    Date &
    Separation $(^{\prime\prime})$ &
    Position Angle $(^\circ)$ &
    RV (m/s)
}
\startdata
1995 Jan 1 & $7.8527 \pm 0.0014$ & $162.844 \pm 0.012$ &     \phn$-1.3 \pm 1.3  $ \\
2000 Jan 1 & $7.4005 \pm 0.0006$ & $165.056 \pm 0.008$ &     \phn$-0.6 \pm 0.5  $ \\
2005 Jan 1 & $6.9029 \pm 0.0026$ & $167.573 \pm 0.022$ & \phs\phn$ 0.5 \pm 0.5  $ \\
2010 Jan 1 & $6.355  \pm 0.008 $ & $170.50  \pm 0.04 $ & \phs\phn$ 2.1 \pm 0.5  $ \\
2015 Jan 1 & $5.748  \pm 0.018 $ & $174.02  \pm 0.07 $ & \phs\phn$ 4.7 \pm 1.9  $ \\
2020 Jan 1 & $5.07   \pm 0.03  $ & $178.42  \pm 0.11 $ & \phs\phn$ 9   \pm 6    $ \\
2025 Jan 1 & $4.31   \pm 0.06  $ & $184.28  \pm 0.21 $ &     \phs$16_{-10}^{+13}$ \\
2030 Jan 1 & $3.45   \pm 0.09  $ & $192.9   \pm 0.5  $ &     \phs$28_{-20}^{+31}$
\enddata
\label{tab:predictions}
\end{deluxetable}

\section{Evolutionary Model Analysis} \label{sec:evol}

For brown dwarfs, only when the fundamental parameters of mass, age and composition are empirically determined, along with some dependent parameter like luminosity, can models be tested. Following our past work \citep[e.g.,][]{2008ApJ...689..436L,2009ApJ...692..729D}, we use the two most precise measurements of mass and luminosity to derive all other parameters from solar-metallicity models (as appropriate for the Gl~229 system). This both provides a test of models, by comparing the model-derived age to the empirical one, and enables more precise (albeit model-dependent) estimates of other parameters like effective temperature (\Teff).

We derive mass-calibrated fundamental parameters for Gl~229B by performing rejection sampling analysis with evolutionary models in the same fashion as in \citet{Dupuy+Liu_2017} and \citet{Dupuy+Liu+Best+etal_2019}. Briefly, we start with randomly drawn masses and luminosities from the measured distributions and combine these with ages randomly drawn from a uniform prior. We use masses drawn from our MCMC posterior and normally distributed values for the luminosity according to the measurement of $\log(\Lbol/\Lsun) = -5.208\pm0.007$\,dex from \citet{Filippazzo2015}. Each random draw corresponds to a measured luminosity as well as a model-derived luminosity (from mass and age), and the rejection probability is computed from the difference between these two luminosities. Over three iterations we adjust the range over which ages are drawn as needed to ensure a well-sampled posterior on model-derived properties. 

\begin{figure}
    \includegraphics[width=0.45\textwidth]{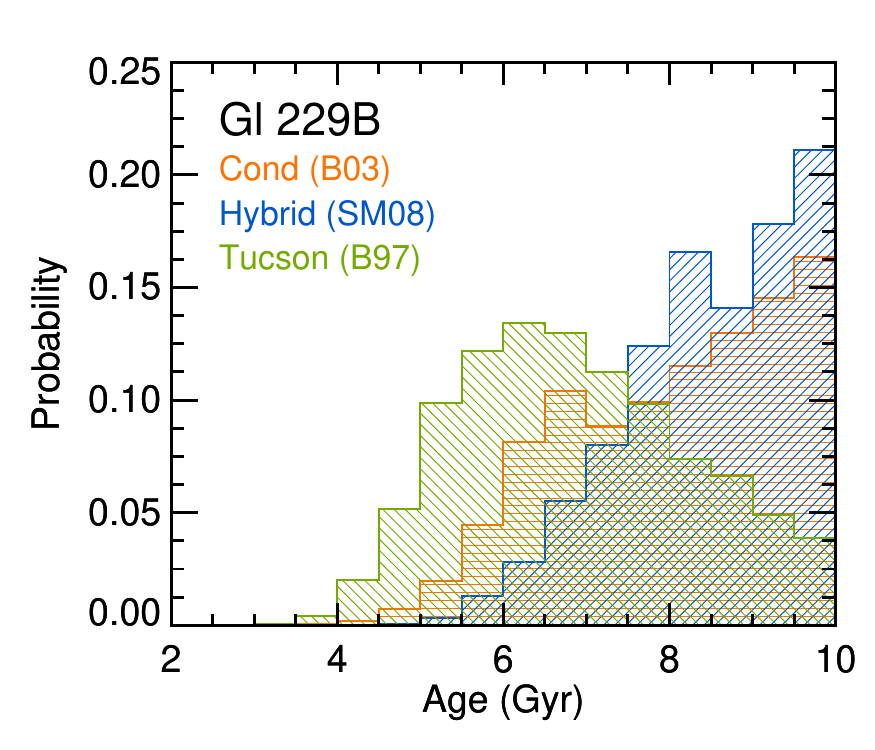}
    \caption{Age distributions derived from substellar evolutionary models given the measured mass and luminosity of Gl~229B. Our rejection sampling analysis assumes a uniform prior on age that, given how broad the output distribution is, strongly influences our result. This ``substellar cooling age'' for the system is somewhat older than implied by the stellar activity of the primary star Gl~229A.} \label{fig:evol-age}
\end{figure}

\begin{deluxetable*}{lcccc}
\tablecaption{Evolutionary Model-Derived Parameters \label{tbl:evol}}
\tablehead{
\colhead{} &
\multicolumn{3}{c}{Model-derived value} &
\colhead{}\\[2pt]
\cline{2-4}\\[-5pt]
\colhead{Property} &
\colhead{Cond (B03)} &
\colhead{Hybrid (SM08)} &
\colhead{Tucson (B97)} &
\colhead{Prior}}
\startdata
Mass (\Mjup)          & $64.5_{-1.1}^{+1.7}$      & $63.1_{-0.9}^{+1.7}$       & $70.1_{-2.1}^{+4.6}$      &      $70\pm5$         \\[3pt]  
log(\Lbol) [\Lsun]    & $-5.208\pm0.007$          & $-5.208\pm0.007$           & $-5.208\pm0.007$          &  $-5.208\pm0.007$     \\[3pt]
Age (Gyr)             & $8.2_{-0.9}^{+1.8}$       & $8.6_{-0.6}^{+1.4}$        & $6.8_{-1.7}^{+1.2}$       & $\mathcal{U}(\leq10)$ \\[3pt]       
\Teff\ (K)            & $1039_{-5}^{+6}$          & $1011_{-5}^{+6}$           & $1023_{-7}^{+9}$          &      \nodata          \\[3pt]
\logg\ [cm\,s$^{-2}$] & $5.451_{-0.014}^{+0.016}$ & $5.400_{-0.010}^{+0.018}$  & $5.467_{-0.019}^{+0.044}$ &      \nodata          \\[3pt]
Radius (\Rjup)        & $0.754_{-0.009}^{+0.003}$ & $0.788_{-0.006}^{+0.003}$  & $0.770_{-0.013}^{+0.006}$ &      \nodata
\enddata 
\tablerefs{B03---\citet{Baraffe+Chabrier+Barman+etal_2003}; B97---\citet{Burrows+Marley+Hubbard+etal_1997}; SM08---\citet{Saumon+Marley_2008}.}
\end{deluxetable*} 

Table~\ref{tbl:evol} gives the results of our evolutionary model analysis, which we performed for three different sets of models: the ``hybrid'' version of the \citet{Saumon+Marley_2008} models, that transitions from cloudy to cloud-free atmospheres as objects cool from 1400\,K to 1200\,K; the ``Cond'' models of \citet{Baraffe+Chabrier+Barman+etal_2003}, that are intended to match the condensate-free atmospheres of T~dwarfs like Gl~229B; and the ``heritage'' Tucson models \citep{Burrows+Marley+Hubbard+etal_1997} that use the cloudless atmospheres of \citet{Marley1996}. Every Monte Carlo trial preserved in our rejection sampling analysis corresponds to a part of parameter space actually covered by models, so the posterior mass of $63.1_{-1.0}^{+1.6}$\,\Mjup\ from hybrid models is significantly lower, with smaller errors, than our input measurement. The modest tension between the posterior mass and the dynamical mass is partially responsible for the small error bar: two incompatible distributions can combine to give a deceptively narrow joint distribution.  The Cond and hybrid models give somewhat different radii (the Cond radius is 4\% smaller), which propagates to gravity and temperature (Cond is higher for both). Even the smaller values of $\Teff  = 1011^{+6}_{-5}$\,K and $\logg = 5.400^{+0.018}_{-0.010}$\,dex from the hybrid models are quite high compared to the typical best-fit model atmospheres in the literature, such as $\Teff = 850$\,K and $\logg = 5.0$\,dex from \citet{Nakajima+Tsuji+Takeda_2015}.  However, these values are consistent with the highest-gravity fits of \cite{Saumon+Geballe+Leggett+etal_2000}.  As noted by \citet{Saumon+Marley_2008}, the Tucson models give a lower luminosity than hybrid models at a given mass and age because they lack many of the opacity sources included in more recent work and $\Lbol \propto \kappa_R^{0.35}$ (where $\kappa_R$ is the Rosseland mean opacity; \citealp{1993RvMP...65..301B}). As a result, the Tucson models provide a younger age estimate ($6.8^{+1.2}_{-1.7}$\,Gyr) and mass posterior closely matching the input measured value for Gl~229B.

\section{Discussion}

Gl~229B provided the first spectrum of a methane-bearing compact source outside our solar system \citep{Nakajima+Oppenheimer+Kulkarni+etal_1995,Oppenheimer+Kulkarni+Matthews+etal_1995} and helped define the ``T'' spectral class \citep{Burgasser1999,Burgasser2002}.  Atmosphere models were developed or refined shortly after Gl~229B's discovery in order to fit its spectrum and estimate its surface gravity, effective temperature, age, and mass \citep{Tsuji1996,Allard1996,Marley1996}. These models suggested a mass of $\sim$20--55~$M_{\rm Jup}$, significantly lower than our dynamical measurement of $70 \pm 5$~$M_{\rm Jup}$. 
Models of T~dwarfs have been tested and refined since then, and molecular line lists have improved \citep[e.g.,][]{2012ApJ...750...74S,2014MNRAS.440.1649Y}.
Although Gl~229B itself lacks an analysis with modern models, the mass range generally implied by spectroscopically-derived gravities for late-T dwarfs has remained broadly similar, consistent with ages of 1--5\,Gyr \citep[e.g.,][]{2017ApJ...848...83L}. 
Because field brown dwarfs are degenerate objects that vary little in radius, gravity is mostly a tracer of mass, and most objects are not observed to have the high surface gravity that would correspond to high mass.

As shown in Figure~\ref{fig:m-l}, Gl~229B has 
one of the highest measured masses of any known brown dwarf
with a luminosity $<3\times10^{-5}$\,\Lsun,\footnote{\cite{Dieterich+Weinberger+Boss+etal_2018} reported a mass of $75.0\pm0.8$\,\Mjup\ for $\epsilon$~Ind~B, which would make it the most massive known T~dwarf. We adopt the mass of $68.0\pm0.9$\,\Mjup\ from \cite{Cardoso_2012} here because, as discussed in \cite{Dupuy+Liu+Best+etal_2019}, it agrees better with the total system mass from the relative orbit measured with VLT/NACO.} though 
within the errors it is consistent with two other massive T~dwarfs $\epsilon$~Ind~B ($68.0\pm0.9$\,\Mjup) and WISE~J0720$-$0846B ($66\pm4$\,\Mjup). 
Figure \ref{fig:Gl229_violin} compares the dynamical mass that we measure to the predictions of the \cite{Saumon+Marley_2008}, \cite{Baraffe+Chabrier+Barman+etal_2003}, and  \cite{Burrows+Marley+Hubbard+etal_1997} model grids.  The left panel assumes a uniform distribution of ages from 1--5~Gyr; the right panel assumes 5--10~Gyr.  At the very oldest ages, the mass is in good agreement with the \cite{Burrows+Marley+Hubbard+etal_1997} models and in marginal agreement with the others.  If the system is younger as suggested by its kinematics and activity indicators, then its dynamical mass of $70\pm 5$~$M_{\rm Jup}$ is surprisingly high for a late-type T~dwarf.

In the following, we examine our unexpectedly high dynamical mass measurement in the context of substellar evolutionary models and the empirical information available on the age of the system. But first, we consider both observational and astrophysical systematic uncertainties that could have impacted our measurement.

\subsection{Potential Systematics}

\subsubsection{Incorrect or Corrupted Host-Star Astrometry}

It is possible that the astrometry for Gl~229A in the {\it Hipparcos} and/or {\it Gaia} DR2 catalogs is incorrect. A slightly high astrometric $\chi^2$ gives some support to this hypothesis (Table~\ref{tab:posteriors}). However, the star passes every goodness-of-fit test in both catalogs, and our orbital fit accounts for the individual scanning epochs and scan angles.  Future \Gaia\ data releases will provide the final word on this possibility.  

It is also possible that another unseen companion is simultaneously perturbing the absolute astrometry and radial velocity of Gl~229A to produce a deceptively good fit at the incorrect mass. For instance, a similar hypothesis was recently invoked to explain the radial velocity trend of HD~206893 \citep{Grandjean+Lagrange+Beust+etal_2019}. In the case here, reducing the mass of Gl~229B by positing an additional companion requires a fortuitous mass and orbit. 
Gl~229A has almost no detectable radial velocity trend ($0.3\pm0.1$\,m\,s$^{-1}$\,yr$^{-1}$ in the HIRES data set, or $0.1 \pm 0.6$\,m\,s$^{-1}$\,yr$^{-1}$ extrapolated to the central epoch of the HGCA, Table \ref{tab:approx_mass}), so an unmodeled companion would have to either be in a nearly face-on orbit or else have a trend that is canceled out by Gl~229B. While we cannot rule out the former, the latter would actually imply an even larger mass for Gl~229B.

A companion on an inner $\sim$decade-long orbit capable of smoothly perturbing Gl~229A's absolute astrometry (but aligned so as to avoid contributing to the radial velocity trend) might not be stable considering Gl~229B's high eccentricity ($\varepsilon = 0.846^{+0.014}_{-0.015}$).  The periastron distance of Gl~229B is $6.1^{+1.1}_{-1.4}$\,AU ($<$12\,AU at 2$\sigma$), making stable orbits unlikely outside $\sim$2\,AU \citep[$\sim$4\,AU at 2$\sigma$,][]{Holman+Wiegert_1999}. This hypothetical inner-orbiting companion would also need to be located on the same side of the star as Gl~229B.  Otherwise, it would cause us to underestimate the acceleration of Gl~229A due to Gl~229B and again imply an even larger mass for Gl~229B.  Such a scenario is difficult to exclude, but requires a finely-tuned mass and orbit for the companion to enable it to evade radial velocity detection.  It also has no supporting evidence beyond a slightly high astrometric $\chi^2$ and the nominal tension with brown dwarf cooling models.

\subsubsection{Unresolved Multiplicity of Gl~229B}

\begin{figure}
    \centering
    \includegraphics[width=\linewidth]{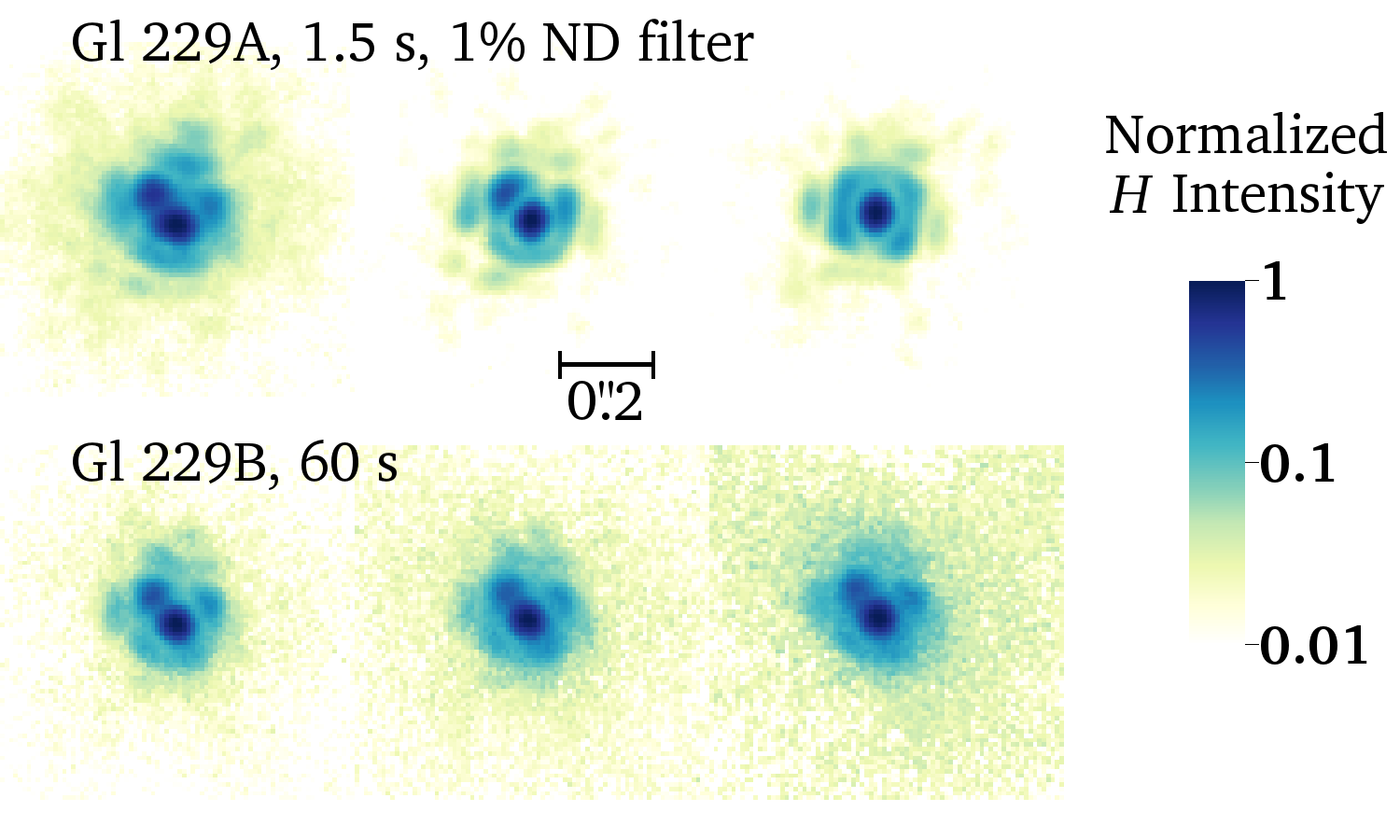}
    \caption{$H$-band images of Gl~229A (top) and Gl~229B (bottom) taken with the HiCIAO camera on the Subaru Telescope.  The images of Gl~229A were taken with a 1\% neutral density filter to prevent saturation.  While the point-spread function varies and is asymmetric, these images show no evidence of Gl~229B itself being a binary. The three lower panels show the only high signal-to-noise images of Gl~229B in the observing sequence.}
    \label{fig:Gl229_PSF}
\end{figure}

One solution to an anomalously high mass is to make Gl~229B itself a binary, e.g., a 45-$M_{\rm Jup}$ object that contributes the vast majority of the observed flux and a much fainter 25-$M_{\rm Jup}$ object.  Evidence of this might appear in high-contrast imaging or in relative astrometry.  Figure \ref{fig:Gl229_PSF} shows the HiCIAO images from 2011: the upper panels show the PSF of Gl~229A, while the lower panels show Gl~229B.  None of these near-infrared images show any evidence for binarity.  The full width at half maximum of the PSF in these cases is $\sim$60~mas, or $\sim$0.3~au at the distance of Gl~229A.

At smaller orbital separations, the precise \HST\ astrometry strongly constrains the presence of additional companions.  The typical separation uncertainties of 1.5~mas project to 
0.009\,AU at the system's distance from Earth, and a typical photocenter orbit for Gl~229B due to a faint companion would be a factor $\sim$0.3 of the semimajor axis. The small $\chi^2$ of 3.0 for our seven relative separation measurements thus provides strong evidence against a massive companion to Gl~229B beyond $a\gtrsim0.03$\,AU ($P\gtrsim7$\,days), up to orbital periods comparable to the time baseline of the measurements ($\approx$5\,years). Even smaller orbital separations, inside 0.01\,AU or 20\,$R_{\rm Jup}$, would have placed Gl~229B in contact with its putative companion when young \citep{Allard+Hauschildt+Alexander+etal_2001}. 
Future measurements with VLT/Gravity, as were obtained for HR~8799e \citep{Gravity_2019}, could offer even better astrometric constraints on a hypothetical companion.

A companion of nearly equal flux and very small separation would evade these astrometric constraints.  However, Gl~229B's absolute magnitude is normal relative to other objects of comparable spectral type and color. For example, the mean T7 dwarf has $M_J = 15.54\pm0.25$\,mag \citep[error is the root-mean-square of four objects][]{Dupuy+Liu_2012} compared to $M_J = 15.21\pm0.05$\,mag for Gl~229B. It is also unremarkable on color--magnitude diagrams, with a very typical absolute magnitude of $M_{L^{\prime}} = 13.44\pm0.05$\,mag at its color of $K-L^{\prime} = 2.12\pm0.07$\,mag \citep[][, Figure~24]{Dupuy+Liu_2012}.

\subsection{Age of Gl~229A}

\begin{figure}
\centering
    \includegraphics[height=1.1\linewidth]{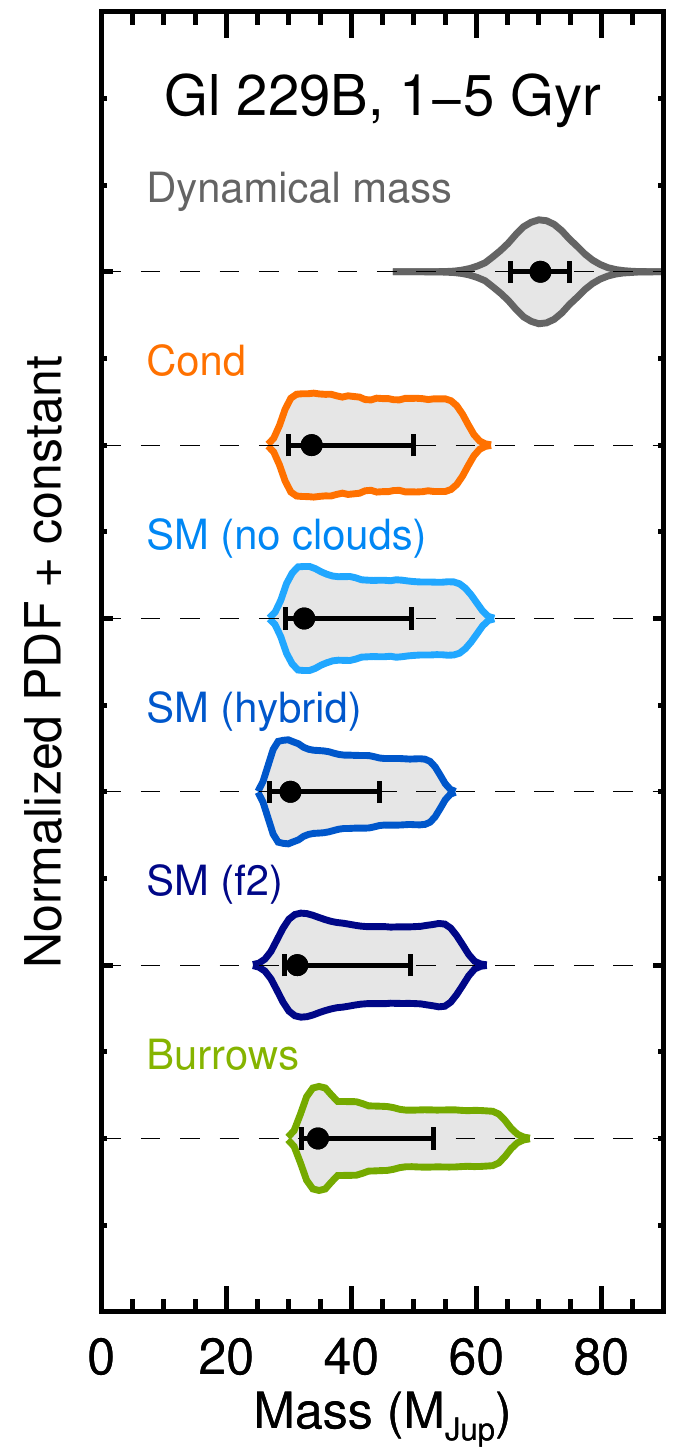}    \includegraphics[height=1.1\linewidth]{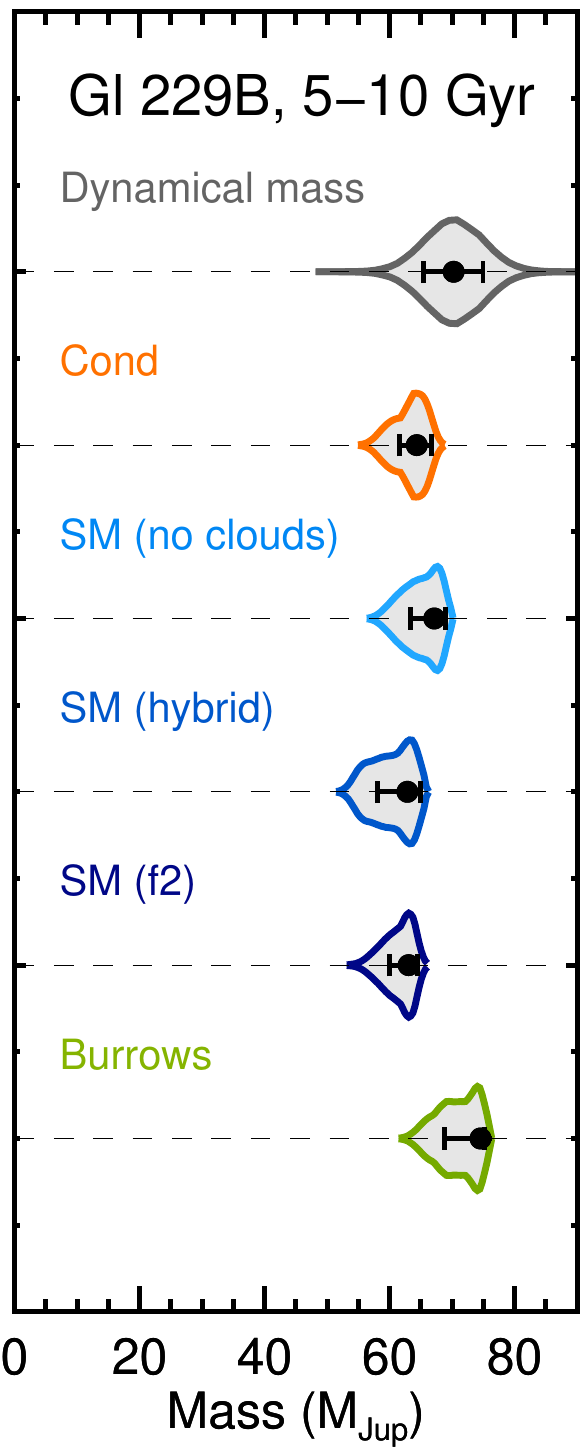}
    \caption{Model-derived masses compared to our dynamical mass of $70\pm 5$~\Mjup\ for two assumed age distributions of Gl~229B: uniform between 1 and 5~Gyr (left panel) and uniform between 5 and 10 Gyr (right panel).  We show the AMES-Cond models \citep{Allard+Hauschildt+Alexander+etal_2001}, three models from \cite{Saumon+Marley_2008}, and the \cite{Burrows+Marley+Hubbard+etal_1997} models.  The models disagree at the intermediate ages suggested by Gl~229A's activity and kinematics (Section \ref{sec:age_metallicity}), but show only modest tension for very old ages. }
    \label{fig:Gl229_violin}
\end{figure}

Gl~229A is an early-M star and does not possess any well-calibrated age indicators. Its chromospheric and coronal activity seem to favor an intermediate age of 2--6\,Gyr, but this relies on modest extrapolations of relations derived for more massive stars.  Its cold, thin-disk kinematics provide another weak constraint disfavoring very old ages.  An age of 7--10~Gyr would bring our mass measurement into agreement with all cooling models (Figure~\ref{fig:Gl229_violin}). Our age analysis (Section \ref{sec:age_metallicity}) disfavors this possibility, but with caveats.

\begin{figure}
    \includegraphics[width=\linewidth]{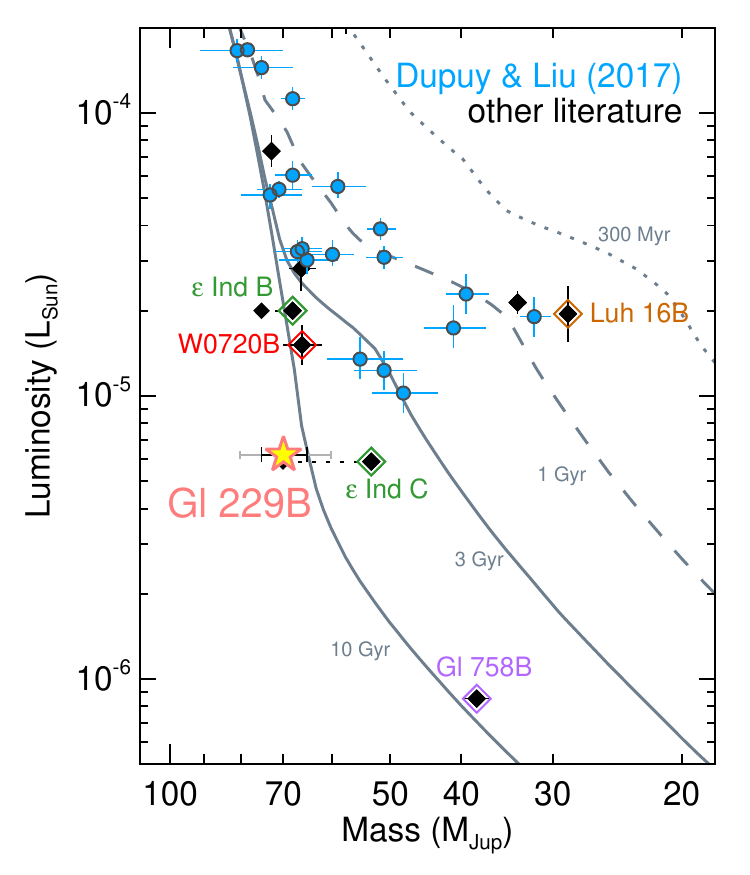}
    \caption{Luminosity as a function of mass for ultracool dwarfs that have model-independent mass measurements, with gray lines showing SM08 hybrid evolutionary model isochrones \citep{Saumon+Marley_2008}. Most measurements come from \citet[][blue circles]{Dupuy+Liu_2017}; other literature measurements are plotted as black diamonds. Notable literature T~dwarfs are highlighted with colored diamonds: $\epsilon$~Ind~B and C (green), Luhman~16B (brown), WISE~J0720$-$0846B (red), and Gl~758B (purple). For $\epsilon$~Ind~BC we show both the lower masses derived by \cite{Cardoso_2012} and the higher masses derived by \cite{Dieterich+Weinberger+Boss+etal_2018}; we connect them with dotted lines. Like $\epsilon$~Ind~B and WISE~J0720$-$0846B our mass measurement for Gl~229B is high for its luminosity; the SM08 models suggest an age of $\sim$10~Gyr.}
    \label{fig:m-l}
\end{figure}

\subsection{Peculiar Spectral Classification}

Although Gl~229B began as the prototypical T~dwarf, once a sufficient sample of spectra for other T~dwarfs became available it was ultimately classified as peculiar \citep{Burgasser2006}. Most T~dwarfs show a consistent sequence of increasing H$_2$O and CH$_4$ absorption across the near-infrared, but Gl~229B has unusually weak absorption at some wavelengths, primarily in the $Y$ and $K$ bands, given the strong CH$_4$ absorption seen at other wavelengths, especially the deep feature on the red side of the $H$ band \citep{1996ApJ...467L.101G,1998ApJ...502..932O}. Early studies proposed that Gl~229B is of subsolar metallicity \citep[e.g.,][]{1998Sci...282.2063G,2000ApJ...532L..59G,Leggett+Hauschildt+Allard+etal_2002}, which is one possible explanation for the spectral peculiarity. There is, however, a well-known degeneracy between metallicity and gravity in T~dwarfs due to collision-induced absorption by molecular hydrogen, the most abundant species in the atmosphere \citep[e.g., see][]{2012ApJ...750...74S}. As noted by \citet{Saumon+Geballe+Leggett+etal_2000}, this degeneracy implies that Gl~229B would have a high surface gravity if it was also of solar metallicity.

As recently pointed out by \citet{Nakajima+Tsuji+Takeda_2015}, previous detailed studies of Gl~229B were done before modern calibrations for empirical M~dwarf abundances were established. While it was once plausible for Gl~229A to have a significantly sub-solar metallicity, that is no longer the case (Section~\ref{subsec:metallicity}). Our dynamical mass measurement suggests that the spectral peculiarity of Gl~229B, compared to other T~dwarfs, is instead 
a reflection of its high surface gravity $\logg = 5.4$--5.5\,dex (Section~\ref{sec:evol}).

\subsection{Reconciling a Young Gl~229 with Brown Dwarf Models}

If we adopt the age of $\sim$3--4~Gyr implied by our kinematic and activity analysis, Gl~229B is significantly less luminous than models predict at its age. These models generally match a fully convective interior to a nongray atmosphere, with varying treatments of molecule and condensate formation, and with different tables of opacities for the various atomic and molecular species \citep{Burrows+Marley+Hubbard+etal_1997,Allard+Hauschildt+Alexander+etal_2001,Saumon+Marley_2008}. Increases in atmospheric opacity tend to increase the observed brown dwarf's radius and thus luminosity \cite{1993RvMP...65..301B}.  A lower predicted luminosity requires either lower opacities or less efficient heat transport out of the core, either at present or at some point in the past.

Both inefficient heat transport out of the core and low opacities go against recent developments in the theory of brown dwarfs. The cores of very massive brown dwarfs like Gl~229B become strongly degenerate at ages $\gtrsim$2\,Gyr; electron conduction can then efficiently transport heat out of the core \citep{2000ApJ...542..464C}.  This increases the predicted luminosity over that of fully convective models, worsening the observed disagreement in Gl~229B, which seems to be at least this old. In addition, predicted mean opacities have generally increased with time as new sources of opacity have been better characterized and line lists have been refined \citep{Freedman+Marley+Lodders_2008,Tennyson+Yurchenko_2012,Morley+Fortney+Marley+etal_2012,Freedman+Lustig-Yaeger+Fortney+etal_2014}, so a significant decrease in predicted luminosity to match Gl~229B's observed properties at a young age would likely require even lower opacities than those of \cite{Burrows+Marley+Hubbard+etal_1997}.  Gl~229B's mass and luminosity are in reasonable agreement with brown dwarf evolutionary models at ages approaching 10\,Gyr (Figures~\ref{fig:Gl229_violin}, and \ref{fig:m-l}). Confirmation of a young age for the Gl~229 system, however, would pose a severe challenge to these models.

\section{Conclusions}

In this paper we have combined radial velocity, imaging, and absolute astrometry to measure a mass of $70 \pm 5$~\Mjup\ for the T dwarf Gl~229B.  We fit Keplerian orbits using an MCMC to derive posterior probability distributions for all of the orbital parameters.  In addition to the high companion mass, we derive an eccentricity of 
$0.846^{+0.014}_{-0.015}$, a semimajor axis of 
$34.7^{+1.3}_{-1.9}$~AU, and an orbital period of 
$263^{+21}_{-29}$~years. Despite the fact that our observations cover only a small fraction of the orbit 
 ($<$10\%), the combination of absolute astrometry and radial velocities establish Gl~229A's three-dimensional acceleration in an inertial reference frame and enable us to determine a precise companion mass.  The uncertainty in our mass measurement is chiefly limited by the proper motion errors of {\it Gaia} DR2; it will improve with upcoming {\it Gaia} data releases. 

Early analyses of the Gl~229 system generally favored an intermediate age and T dwarf companion mass of $\sim$20--55~\Mjup\ \citep{Nakajima+Oppenheimer+Kulkarni+etal_1995,Oppenheimer+Kulkarni+Matthews+etal_1995,Leggett+Toomey+Geballe+etal_1999,Leggett+Hauschildt+Allard+etal_2002}.  Evolutionary models are marginally or fully compatible with the higher mass that we measure if the system is old 
(7--10~Gyr).  We revisit Gl~229A's age using both kinematics and stellar activity.  Both disfavor such old ages, but with caveats: kinematics provide only limited precision while the activity relations must be extrapolated from mid-K to early-M spectral types.  If the brown dwarf cooling models are reliable, then the dynamical mass that we measure could instead be used to calibrate stellar activity relations for M~dwarfs at very old ages.

Given the disagreement between our measurement and earlier mass estimates, we address factors that could inflate our measured mass of Gl~229B.  Our radial velocity time series comes from the HIRES \'echelle spectrograph on Keck as released by \cite{Butler+Vogt+Laughlin+etal_2017}; it shows no evidence of the inner planet reported by \cite{Tuomi+Jones+Barnes+etal_2014}. Positing another massive companion would induce a trend that Gl~229B would have to cancel, further increasing its inferred mass.  Relative astrometry also disfavors another massive companion in the system,
as it would perturb the archival {\it HST} imaging that we re-analyzed to achieve uncertainties of 1.5~mas.
High-contrast imaging using HiCIAO on the Subaru Telescope shows no evidence for a luminous companion beyond $\gtrsim$0.3\,AU.  While the $\chi^2$ of the absolute astrometry is somewhat high, at 7.4 for roughly three degrees of freedom, Gl~229A is well-fit as a single star in both {\it Hipparcos} and {\it Gaia}.  Future {\it Gaia} data releases will provide a definitive answer on the absolute astrometry. 
Overall, we conclude that our measured mass should accurately reflect the true mass of Gl~229B, and while this is somewhat higher than expected, it is consistent with Gl~229B being an old and commensurately massive T~dwarf.

With our mass measurement, Gl~229B joins a short but growing list of ultracool dwarfs just below stellar/substellar boundary. 
Gl~229B is the lowest-luminosity brown dwarf known to have such a high mass, thereby providing a critical test of substellar evolution theory.

\acknowledgements{This research is based in part on data collected at the Subaru telescope, which is operated by the National Astronomical Observatories of Japan. The authors wish to recognize and acknowledge the very significant cultural role and reverence that the summit of Maunakea has always had within the indigenous Hawaiian community.  We are most fortunate to have the opportunity to conduct observations from this mountain.  This work has made use of data from the European Space Agency (ESA) mission {\it Gaia} (https://www.cosmos.esa.int/gaia), processed by the Gaia Data Processing and Analysis Consortium (DPAC, https://www.cosmos.esa.int/web/gaia/dpac/consortium). Funding for the DPAC has been provided by national institutions, in particular the institutions participating in the {\it Gaia} Multilateral Agreement.  T.D.B.~and J.F.~gratefully acknowledge support from the Heising-Simons foundation.  
T.D.B.~acknowledges support from NASA under grant \#80NSSC18K0439. T.J.D.~acknowledges research support from Gemini Observatory, which is operated by the Association of Universities for Research in Astronomy, Inc., on behalf of the international Gemini partnership of Argentina, Brazil, Canada, Chile, the Republic of Korea, and the United States of America. B.P.B.~acknowledges support from the National Science Foundation grant AST-1909209. D.M.~gratefully acknowledges support from the European Space Agency (ESA) Research Fellow programme.
This work benefited from the 2019 Exoplanet Summer Program in the Other Worlds Laboratory (OWL) at the University of California, Santa Cruz, a program funded by the Heising-Simons Foundation.}

\bibliographystyle{apj_eprint}
\bibliography{refs,refs-0720,tdupuy}

\end{document}